\documentclass[fleqn,usenatbib]{mnras}

\usepackage{mathptmx}
\usepackage{txfonts}
\usepackage{graphicx}
\usepackage{epstopdf}
\usepackage{caption}
\usepackage{subcaption}
\usepackage{array}
\usepackage{hyperref}
\usepackage{makecell}
\usepackage[normalem]{ulem}
\usepackage{xcolor}

\usepackage[T1]{fontenc}

\DeclareRobustCommand{\VAN}[3]{#2}
\let\VANthebibliography\thebibliography
\def\thebibliography{\DeclareRobustCommand{\VAN}[3]{##3}\VANthebibliography}

\newcolumntype{P}[1]{>{\centering\arraybackslash}p{#1}}

\title[Massive star-formation in RCW~117]{Massive star-formation in the hub-filament system of RCW~117}

\author[Arun Seshadri et al.]{
Arun Seshadri,$^{1}$\thanks{E-mail: arun.seshadri95@gmail.com}
S. Vig,$^{1}$
S.K. Ghosh$^{2}$
and D.K. Ojha$^{2}$
\\
$^{1}$Indian Institute of Space science and Technology, Thiruvananthapuram - 695547, India\\
$^{2}$Tata Institute of Fundamental Research (TIFR), Mumbai, India\\
}

\date{Accepted XXX. Received YYY; in original form ZZZ}

\pubyear{2023}

\begin{document}
\label{firstpage}
\pagerange{\pageref{firstpage}--\pageref{lastpage}}
\maketitle

\begin{abstract}
We present a multiwavelength investigation of the hub-filament system RCW~117 (IRAS~17059-4132), which shows intricate filamentary features in the far-infrared, mapped using \textit{Herschel} images. We obtain the column density and dust temperature maps for the region using the \textit{Herschel} images, and identify 88 cores and 12 filaments from the column density map of the region ($18'\times18'$). We employ the ThrUMMS $^{13}$CO (J=1-0) data for probing the kinematics in RCW~117, and find velocity gradients ($\sim 0.3-1$~km~s$^{-1}$~pc$^{-1}$) with hints of matter inflow along the filamentary structures. Ionised gas emission from the associated HII region is examined using the Giant Metrewave Radio Telescope (GMRT) at 610 and 1280~MHz, and is found to be of extent $5 \times 3$~pc$^2$ with intensity being brightest towards the hub. We estimate the peak electron density towards the hub to be $\sim 750$~cm$^{-3}$. Thirty four Class~0/I young stellar objects (YSOs) have been identified in the region using the \textit{Spitzer} GLIMPSE colour-colour diagram, with many lying along the filamentary structures. Based on the (i) presence of filamentary structures, (ii) distribution of cores across the region, with $\sim39$\% found along the filamentary structures, (iii) massive star-formation tracers in the hub, and (iv) the kinematics, we believe that global hierarchical collapse can plausibly explain the observed features in RCW~117.
\end{abstract}




\section{Introduction}
 
The formation of massive stars (M$\gtrsim8$~M$_\odot$) is an arena of active research as relatively less is known about their formation mechanism compared to their lower mass brethren  \citep{2007ARA&A..45..481Z, 2015IAUGA..2254046T, 2018ARA&A..56...41M}. As massive stars are rare and spend upto 15\% of their lifetimes in the embedded phase within dense molecular clouds \citep{2002ARA&A..40...27C}, infrared and longer wavelengths are ideal tools for their study. \citet{2018ARA&A..56...41M} give a possible evolutionary sequence for high-mass stars - Infrared Dark Clouds (IRDCs), massive starless clumps, high-mass prestellar cores, massive protostars, and hot molecular cores. The later stage involves the development of  hypercompact or ultracompact H{\small II} (HC/UCH{\small II}) regions, which evolve further into classical H{\small II} regions and act as radio beacons for young massive stars. Other conspicuous effects of massive star-birth include jets and outflows that observationally manifest as motion of gas, fluoroscence and shock-excitation in regions of the surrounding medium \citep{2014prpl.conf..451F, 2016ARA&A..54..491B}, as well as the presence of masers \citep[][and references therein]{2002A&A...390..289B}. It has also been observed that massive stars do not form in isolation, but are formed accompanied by low mass stars \citep{2003ARA&A..41...57L, 2007ARA&A..45..481Z, 2009MNRAS.400.1775S}.

Observations of dense clouds of gas using molecular line and far-infrared observations have revealed their highly filamentary in nature \citep{1984ApJ...278L..19L, 2010A&A...518L.103M, 2010A&A...518L.102A, 2013A&A...553A.119A, 2013A&A...554L...2Z}. 
IRDCs are dense, cold regions appearing dark against the bright infrared Galactic background, which are  believed to be the birthplaces of massive stars and clusters \citep{2006ApJ...641..389R, 2009ApJS..181..360C}. Most IRDCs display elongated morphology, with sizes varying from a few  parsecs to hundreds of parsecs \citep{2010ApJ...719L.185J, 2014A&A...568A..73R}. Yet another feature that is encountered in infrared and submillimeter observations is the hub-filamentary structure, characterised by the presence of a central hub onto which filaments appear to converge radially. Higher resolution studies of the dark clouds with \textit{Herschel Space Observatory} and Atacama Large Millimetre Array (ALMA) have shown that hub structures tend to form cluster of stars often containing massive protostars \citep{2009ApJ...700.1609M, 2014A&A...561A..83P, 2019A&A...629A..81T, 2021A&A...646A.137L, 2022A&A...658A.114K}. The central hubs are found to have relatively high column densities of $\sim10^{23}$~cm\textsuperscript{-2} as compared to the filaments, and hence undergo collapse and fragmentation, resulting in cluster formation. Accretion of material via filaments to the clumps and cores in the hub have been detected through molecular line observations \citep{2012ApJ...756...10L, 2018A&A...610A..77H, 2020ApJ...905..158W}. The formation of filaments, and their characteristics such as widths, linear mass densities, etc are found to be related to the properties of the interstellar medium such as gravity and turbulence \citep{2014prpl.conf...27A}.  

\begin{figure*}
	\includegraphics[width=\textwidth]{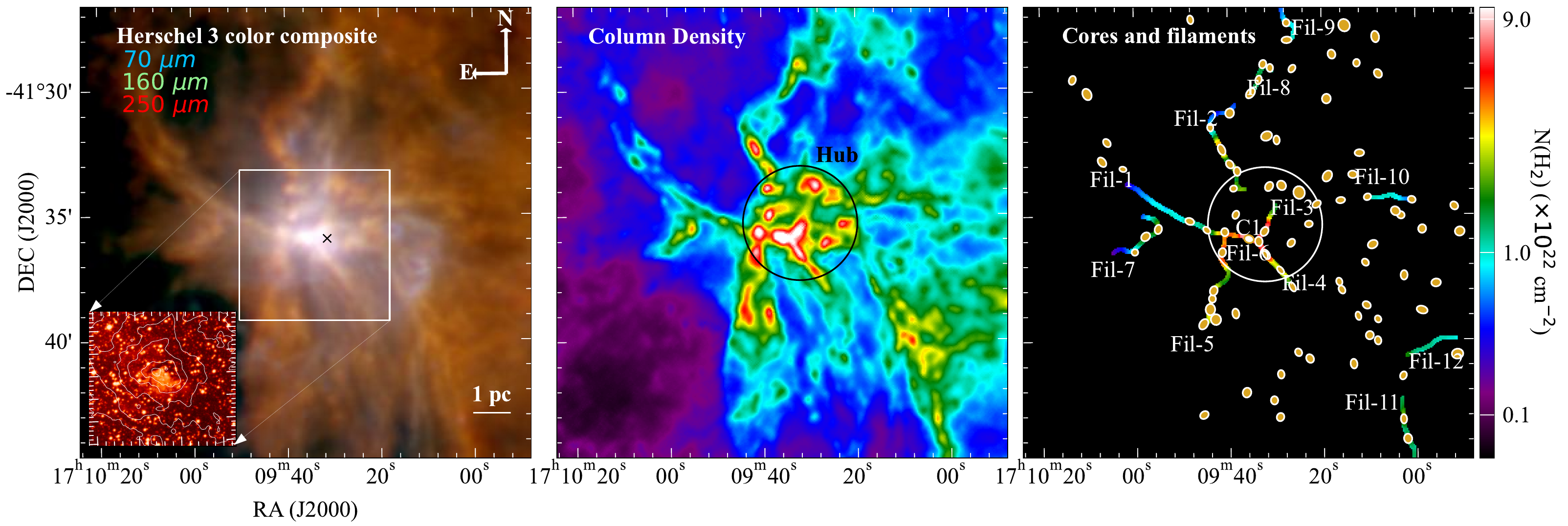}
	\caption{(a) \textit{Herschel} colour-composite image of RCW~117 comprising of emission at 350~$\mu$m (red), 160~$\mu$m (green) and 70~$\mu$m (blue), showing the hub filamentary structure. The IRAS object 17059-4132 is marked as a black cross. (Inset) Rectangular region shown in the colour-composite image, showing the optical nebula associated with RCW~117 from the Digitized Sky Survey (DSS). The contours represent the 70~$\mu$m emission, with levels 0.25, 0.75, 2.5, 7.5, 20, 75 Jy/pixel. (b) The column density map for RCW~117 is shown. The central hub is marked as a black circle. (c) Cores and filament skeletons extracted using \textit{getsf}. The white solid circle encloses the hub. The most massive core, C1, is marked using a white solid ellipse within the hub.}
	\label{fig:herchel_complete}
\end{figure*}

\begin{table*}
	\centering
	\caption{Details regarding the archival datasets used in this work.}
	\label{tab:archdata}
	\begin{tabular}{P{3cm}P{3cm}P{3cm}P{3cm}P{3cm}} 
		\hline \hline
		Survey name, Telescope & Wavelength & Resolution ($''$) & Pixel size ($''$) & Reference  \\
		\hline
		\makecell{GLIMPSE,\\ \textit{Spitzer} Space Telescope} & 3.6, 4.5, 5.8, 8.0 $\mu$m & 2 & 0.6 & \cite{2004ApJS..154....1W} \\
        \\
		\makecell{Hi-Gal,\\ \textit{Herschel} Space Telecope} & 70, 160, 250, 350, 500 $\mu$m & 8.4, 13.4, 18.2, 24.3, 36 & 3.2, 3.2, 6, 10, 14 & \cite{2010PASP..122..314M} \\
		\\
        \makecell{ThrUMMS,\\ 22m Mopra single dish} & $^{13}$CO (J=1-0) (2.6 mm) & 72 & 24 & \cite{Barnes_2015} \\
		\\
        \makecell{MALT90,\\ 22m Mopra single dish} & 3.3mm & 38 & 9 & \citet{2013PASA...30...57J, 2013PASA...30...38F}\\

		\hline \hline
	\end{tabular}

\end{table*}


\begin{table}
	\centering
	\caption{Details of radio continuum observations carried out with the GMRT, India.}
	\label{tab:table2}
	\begin{tabular}{ccc} 
		\hline \hline
		Frequency & 610 MHz & 1280 MHz \\
		\hline
		Observation date & 9 April 2004  &  2 August 2003\\
		On-source integration time (Hrs) & $\sim$1.8 & $\sim$1.4\\
		Bandwidth & 16 MHz & 16 MHz \\
		Flux calibrators & 3C286 & 3C286 \\
		Phase calibrators & 1822-096 & 1626-298 \\
		Synthesized beam & $12.3''\times 5.6''$ & $5.9''\times 2.4'' $\\
		Position angle (deg) & -17.8 & -19.6\\
		Noise (mJy/beam) & 5.4 & 4.2\\
		Convolved beam & $13''\times13''$ & $6''\times6''$\\
		\hline \hline
	\end{tabular}
\end{table}


In the present study, we investigate the region associated with the optical nebula RCW~117 \citep{1960MNRAS.121..103R}, which also hosts the IRAS object 17059-4132 (hereafter I17059), classified as an H{\small II} region. This region lies towards Scorpius about a degree away from the Galactic mid-plane in the fourth Galactic quadrant towards the longitudinal direction $\sim345^{\circ}$. Kinematic distance estimates of RCW~117 in literature range between 1.8 and 2.5~kpc \citep{2003A&A...405..639P, 2004AJ....128.2374S, 2014ApJS..212....2G}. In this work, we adopt a distance of 2.5 kpc for this region, considering LSR~velocity of $\sim-21$~km~s$^{-1}$ \citep{2015A&A...579A..91W} and the Galactic rotation parameters from \citet{2014ApJ...783..130R}. \citet{2004AJ....128.2374S} have provided distribution of C$^{18}$O and HC\textsubscript{3}N maps towards this region, and they conclude that the cloud is likely to be centrally condensed and gravitationally bound. A number of masers have been detected towards this region, such as 	CH\textsubscript{3}OH, H\textsubscript{2}O, and OH \citep{1995MNRAS.272...96C, 1995MNRAS.274..808C, 2003MNRAS.341..551C}.

In the current work, we demonstrate that this region forms an interesting case study for hub-filamentary systems as the dust emission at far-infrared reveals numerous filaments converging to a hub. Our motivation in this work is to characterise the filaments as well as the hub structure, and to study the large scale kinematics of gas. We probe the distribution of ionised gas, molecular gas and dust emission to probe the star-formation activity. In particular, we study the hub that ensconces the massive stars that ionise the gas in detail. The organisation of the paper is as follows. In Sect.~\ref{sec:data}, we present the details of the data used in the paper. The results are given in Sect.~\ref{sec:results}, and discussed in Sect.~\ref{sec:discussions}. Finally, the conclusions of the work are presented in Sect.~\ref{sec:conclusions}.

\section{Data used} \label{sec:data}

We have utilised multi-wavelength data at infrared and millimetre wavelengths from a number of archives to probe the gas and dust associated with the region RCW~117, the details of which are given in Table~\ref{tab:archdata}. The data from Hi-Gal survey, implemented using the \textit{Herschel} Space Observatory is used for mapping the cold dust emission in the region. The gas kinematics in the region are studied using the molecular line emission obtained from \textit{Three Millimeter Ultimate Mopra Survey} (ThrUMMS) and \textit{Millimeter Astronomy Legacy Team 90 GHz} (MALT90) surveys. The embedded young stellar objects (YSOs) in the region are identified using \textit{Spitzer Galactic Legacy Infrared Midplane Survey Extraordinaire} (GLIMPSE).

In addition, ionized gas emission at 1280 and 610~MHz towards this region has been observed using the Giant Metrewave Radio Telescope \citep[GMRT; ][]{1991CSci...60...95S}. GMRT has a total of 30 operational antennas, each of diameter of 45~m. Twelve antennas are distributed randomly in a central array of size $\sim1$~km$^2$, while the remaining 18 antennas are spread out along three arms in a Y-shaped configuration. There are 6 antennas along each arm of the Y that extends upto $\sim14$~km. 
 
 The largest structures that can be mapped at 610 MHz and 1280 MHz are $17'$ and $8'$, respectively. The flux calibrator used for both observations was 3C286, while the phase calibrators used were 1822-096 and 1626-298 at 610 and 1280 MHz, respectively. The details of the observations are given in Table~\ref{tab:table2}. 

The data reduction was carried out using the NRAO Astronomical Image Processing System (\texttt{AIPS}). The tasks \texttt{TVFLG} and \texttt{UVFLG} were used for flagging visibilities corrupted due to non-working antennas and radio frequency interference. The data were then calibrated and deconvolved using the task \texttt{IMAGR}. A few rounds of self-calibration were carried out during imaging to correct for amplitude and phase errors, and the final image was obtained. The flux densities of the target source are rescaled to account for the change in system temperature due to sky noise from the Galactic plane as (a)
the flux calibrator is located away from the Galactic plane, and (b) the observations
 were carried out with the Automatic Level Corrector (ALC) off. The rescaling factor was calculated using an improved
 version \citep{2015MNRAS.451.4311R} of the continuum all-sky map of \citet{1982A&AS...47....1H} at 408 MHz, and assuming a spectral index of -2.6 \citep{1999A&AS..137....7R, 2011A&A...525A.138G}. The effect of the primary beam response was removed using the task \texttt{PBCOR}. The rms noise and the synthesized beams obtained in the final images are listed in Table~\ref{tab:table2}. Finally, the images were convolved to circular beams of sizes $13''$ and $6''$ at 610 and 1280~MHz images, respectively, in order to compensate for the high ellipticity of beams caused due to the relatively low and skewed \textit{uv}  coverage. 
 
\section{Results} \label{sec:results}

\subsection{Cold dust emission}

The cloud associated with RCW~117 is traced using far-infrared cold dust emission from \textit{Herschel} Hi-Gal \citep{2010A&A...518L.100M}, using the longer wavebands $\ge160$~$\mu$m. These images, shown in Fig.~\ref{fig:herchel_complete}(a), depict numerous filamentary structures that appear to converge towards a central region or the hub; with I17059 (error ellipse size $\sim18''\times6''$) lying with in it. The extent of the large scale hub-filamentary structure is $\sim18'\sim13.1$~pc. The filaments appear fragmented, and the longest one extends to about $4'\sim 3$~pc. The emission at 70~$\mu$m closely resembles emission at mid-infrared (MIR) bands. The optical nebula of RCW~117 of size about $1.5'\times1'$ \citep{1960MNRAS.121..103R} lies close to the hub and shows elongated features in absorption, see inset in Fig.~\ref{fig:herchel_complete}(a). 

 To characterise the morphological structures in the \textit{Herschel} Hi-Gal images (160, 250, 350 and 500~$\mu$m), we have employed the \textit{getsf} software package that is capable of extracting cores and filaments by separating their structural components in the images \citep[]{2021A&A...649A..89M, 2021A&A...654A..78M}. For this, we have utilised the images at their native resolutions (listed in Table~\ref{tab:archdata}). The 250~$\mu$m image at native resolution has a few saturated pixels towards the hub. These were replaced by values from an image generated by convolving the high resolution map ($\sim9''$) at 250~$\mu$m (that was available as a highly processed image from the Herschel Science Archive) to the native resolution. After resampling and regridding the multiwavelength images to the resolution of the 160~$\mu$m map, a pixel-to-pixel fit using a modified blackbody function for optically thin emission was carried out using the \textit{hires} package in the \textit{getsf} software suite; the details of which are given in \citet{2016A&A...593A..71M, 2021A&A...654A..78M}. The output includes column density and dust temperature maps at a resolution of $13.5''$. 

The column density map is shown in Fig.~\ref{fig:herchel_complete}(b). The column densities of the cloud span two orders of magnitude from  $\sim 5\times10^{21}$~cm$^{-2}$ to $2\times10^{23}$~cm$^{-2}$ with the highest column density ($2.2\times10^{23}$~cm$^{-2}$) located towards the centre of the hub. Based on the column density image and the $^{13}$CO integrated intensity map (see Sec. \ref{CO}), we take the hub as a circular region of diameter $2.3'$ centered at $\alpha_{\textrm{J}2000}=17^{h}09^{m}32.3^{s}$,  $\delta_{\textrm{J}2000}=-41^{\circ}35'22.1''$, shown in Fig.~\ref{fig:herchel_complete}(b). The temperature of the cold dust lies in the range $\sim 20 - 40$~K, with the hotter dust located towards the hub. These values are found to be in agreement with those of other hub-filament systems in literature. For instance, towards the hub in the hub-filament system of Monoceros~R2, \citet{2015A&A...584A...4D} find $N(H_{2}) \sim  2 \times 10^{23}$~cm$^{-2}$ and $T_{dust} \sim 27$~K. \cite{2022ApJ...934....2M} have investigated the hub-filament systems in the W31 region, and find column density values to be $\sim10^{22}$ - 10$^{23}$ cm$^{-2}$. The corresponding dust temperature towards the hubs is found to vary between $21-32$~K. \cite{2022ApJ...931..115W} find the maximum column density of $10^{23}$~cm$^{-2}$ towards the hub-filament system SDC13.


\begin{figure*}
	\includegraphics[width=\textwidth]{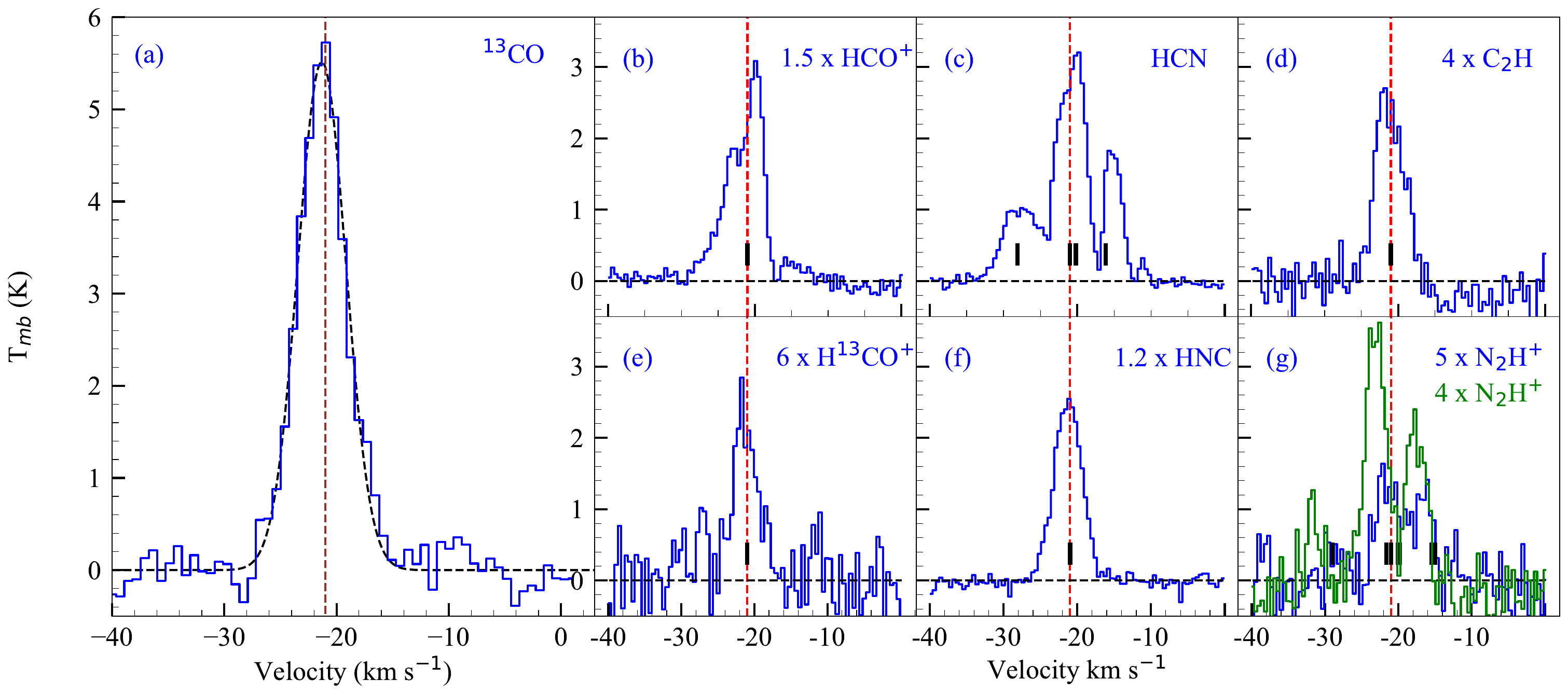}
	\caption{(a) ThrUMMS spectrum of $^{13}$CO (J=1-0) towards the hub is shown in blue. The best-fit model to the observed spectrum is shown as the black dashed-line, more details in text. The vertical red dashed line represents the LSR velocity. (b) - (g) MALT90 spectra of molecules integrated towards a circular region in the hub (shown as blue circle in Fig.~\ref{fig:malt90mom0}), and marked in blue. The spectrum marked in green N$_2$H$^+$ is extracted from a region distinguished by a green circle in Fig.~\ref{fig:malt90mom0}.
 }
	 \label{fig:spectra}
\end{figure*}  


\begin{figure*}
	\includegraphics[width=\textwidth]{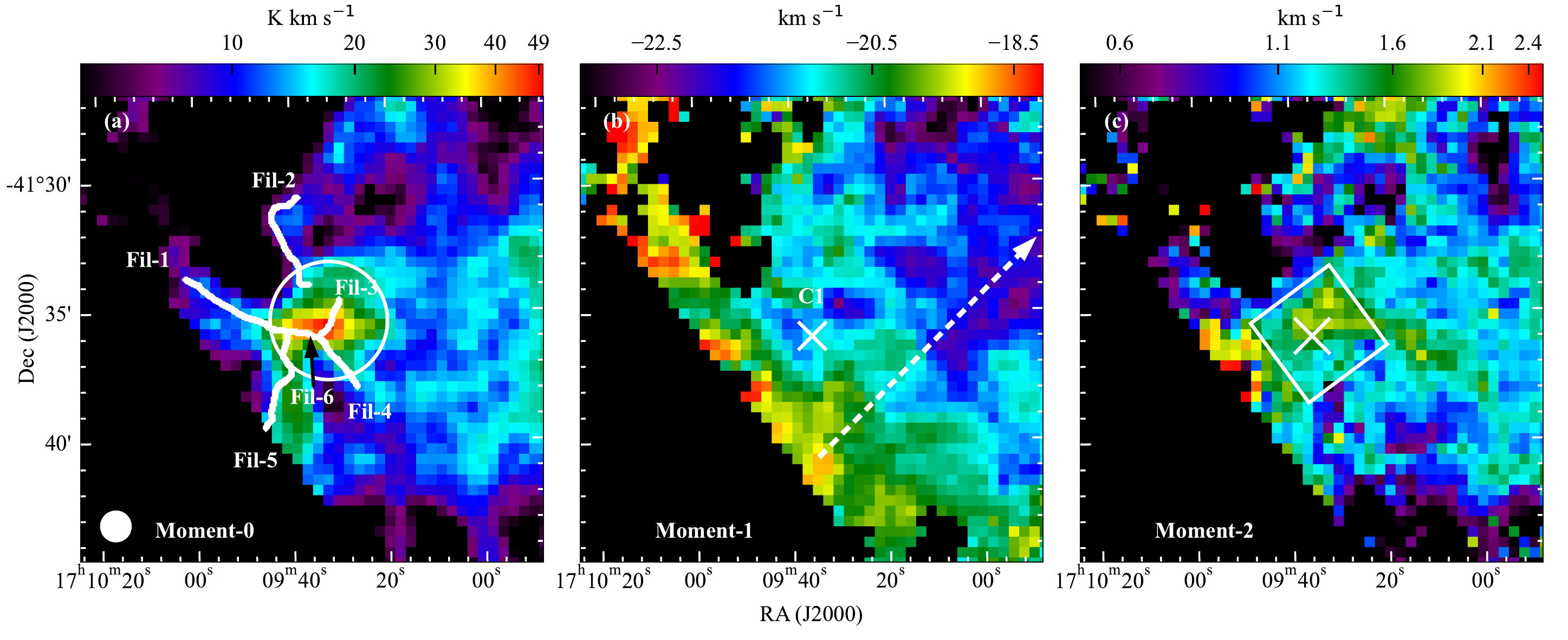}
	\caption{Moment maps generated from the ThrUMMS $^{13}$CO~($J=1-0$) data. (a) Integrated intensity (Moment-0) map integrated  over a velocity range of -27~km~s$^{-1}$ to -15~km~s$^{-1}$. The circle marks the hub while the filamentary structures towards the hub and within are shown. (b) Intensity weighted velocity (Moment-1) map. The most massive core C1 is marked as a white cross, while the arrow indicates direction of the global velocity gradient. (c) Intensity weighted dispersion (Moment-2) map. The white square displays the coverage of MALT90 images. }
	\label{fig:co_moments}
\end{figure*} 


The column density image is employed for the identification and extraction of filaments, and localised over-densities designated as cores; these are shown in Fig.~\ref{fig:herchel_complete}(c). The detailed description regarding the extraction and measurement procedures can be found in \cite{2021A&A...649A..89M} and references therein. A total of 88 cores are extracted in the region under consideration, shown in Fig.~\ref{fig:herchel_complete}(c). Of these, 14 cores are located within the hub, and the brightest (most massive) core in the hub is designated as C1. The other cores are labelled in the order of decreasing mass, which are tabulated in Table~\ref{tab:core_properties} in Appendix A. The mean core temperatures lie in the range $\sim16-37$~K, while the peak column densities are in the range $\sim 6 \times 10^{21}$ - $2 \times 10^{23}$~cm$^{-2}$. The deconvolved effective sizes of the cores are estimated 
using the expression $D_{eff} = \sqrt{(D_{core}^{2}-\theta_{beam}^{2})}$, where $D_{core}$ represents the geometric mean of the major and minor core diameters and $\theta_{beam}$ corresponds to the spatial resolution at 2.5~kpc. The core sizes are found to be in the range $0.07 - 0.3$~pc. These values are in broad agreement with other studies in literature, where core temperatures and radii are found to lie in the range $14-45$~K and $0.01-0.16$~pc, respectively \citep{{2021MNRAS.508.2964A}, {2022MNRAS.514.6038Z}}.

A total of 12 filaments are detected in the region with an aspect ratio (length-to-width ratio) of 3 or greater, similar to \cite{2019A&A...621A..42A}. Towards the hub, we note two locations where filaments appear to converge onto a bridge filament Fil-6, with C1 located towards the western edge of Fil-6. Outside the hub, we observe filaments extending to the north, east, west and southern directions. We find the average filament temperatures to be in the range $16.5-29.5$~K, and the average column density of filaments to be $\sim10^{22}$~cm$^{-2}$. This is consistent with the values obtained in other studies \citep{2010A&A...518L.100M, 2019A&A...621A..42A, 2019A&A...629A..81T}. We note that these values are lower than the corresponding values of cores. A comparison of distribution of cores and filaments shows that 34 cores ($\sim39$\%) lie along the detected filaments suggesting that the filaments are undergoing fragmentation. We discuss this aspect further in Sect.~\ref{disc:scenario}.

We have estimated the filament widths by deconvolving a Gaussian beam corresponding to the resolution of \textit{Herschel} 160~$\mu$m image under a simplistic assumption that the filament widths have Gaussian profiles. We find the deconvolved filament widths of this region to lie in the range $0.3-0.5$~pc. These widths are larger than the characteristic widths of $\sim0.1$~pc found for clouds in the solar neighbourhood \citep{2011A&A...529L...6A, 2019A&A...621A..42A, 2022A&A...667L...1A}. The extraction of actual filament widths from the images using deconvolution techniques could be distance dependant \citep{2022A&A...657L..13P, 2022A&A...667L...1A} and further in-depth analyses are required to establish this firmly. We, therefore, do not proceed further with the analysis of filament widths. 

The masses of the cores are estimated by means of the equation: $M=\mu m_{H} N(H_{2})A$, where $\mu=2.86$ is the mean molecular weight of the particles in the cloud, assuming that 70\% of the gas is H$_2$, $m_{H}$ is the mass of the hydrogen atom, $N(H_{2})$ is the integrated column density of the core or filament, and $A$ is the pixel area in cm$^2$.  The cores have masses in the range $1-174$~M$_{\odot}$, with C1 being the most massive core. The cores located within the hub are relatively more massive than the ones located outside, with masses of the former in the range $\sim60 - 174$~M$_{\odot}$. The masses of cores are consistent with those found towards other regions; in the range $1-1000$~M$_{\odot}$ \citep{2012A&A...541A..12J, 2021MNRAS.508.2964A}. The total mass within the hub as estimated from the column density map is $\sim5700$~M$_{\odot}$.

\begin{figure*}
	\includegraphics[width=0.95\textwidth]{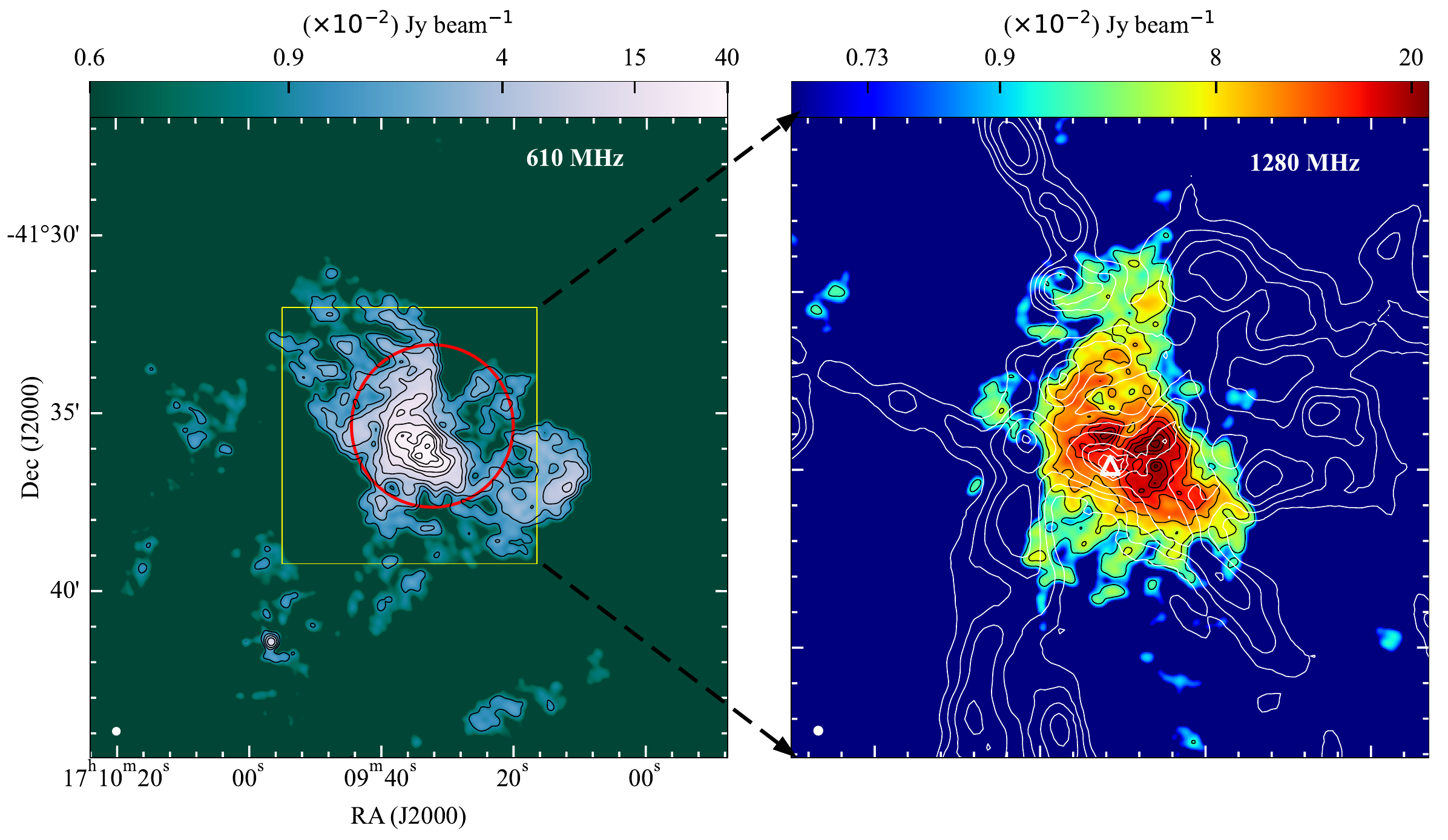}
	\caption{(Left) 610~MHz emission towards RCW~117 observed using GMRT. The contours levels are at 0.01, 0.03, 0.09, 0.2, 0.3, and 0.395~Jy/beam, where the beam $\sim 13''\times13''$, also shown at the bottom left corner of the image. The hub is marked using a red circle. (Right) 1280~MHz emission observed with GMRT, with levels at 0.0018, 0.018, 0.075, 0.18, 0.35~Jy/beam, where the beam is $\sim 6''\times6''$. The contours represent column density with levels at $1.3\times~10^{22}$, $2\times~10^{22}$, $3\times~10^{22}$, $5\times~10^{22}$, $10\times~10^{22}$, $13\times~10^{22}$, $16\times~10^{22}$ cm$^{-2}$. The white triangle indicates the most massive core C1.}
	\label{fig:radiopanel}
\end{figure*}  


\subsection{Warm dust emission} \label{sec:warmdust}

The mid-infrared (MIR) emission towards RCW~117 with a spatial extent of $\sim 2.6\times 3.0$~pc$^2$, is probed using \textit{Spitzer}-GLIMPSE images between 3.6 and 8~$\mu$m. This emission is primarily attributed to stochastically heated small dust grains \citep{2003ARA&A..41..241D} and Polycyclic Aromatic Hydrocarbon (PAH) molecules present in the photodissociation regions (PDRs) \citep{2015ApJ...807...99A}. We note numerous filamentary and arc-like features in the emission, which suggests a flurry of star-formation activity. It has been reported that the IRAC 4.5 $\mu$m band can trace shocked gas through emission in H$_2$ and CO lines, indicative of protostellar outflows \citep{2008MNRAS.384...71T, Ray2023}. We have constructed the [4.5]/[3.6] $\mu$m flux ratio map in order to localise regions of shocked gas. A ratio of $\geq 1.5$ is a signature of the presence of shocked gas, while $\ll 1.5$ is indicative of stellar sources \citep{2004ApJS..154..352N, 2010ApJ...720..155T}. Towards RCW~117, we find a few regions having ratio $\geq 1.5$, which lie in the vicinity of C1 towards the centre of the hub, again demonstrating the turmoil in the ISM caused due to star-formation here.

\subsection{Molecular line emission}
We have utilised ThrUMMS and MALT90 surveys in order to understand the kinematics of the gas in the region. While the ThrUMMS survey has a relatively large coverage of our region of interest towards RCW~117, the MALT90 survey has a smaller coverage of $4'\times4'$ towards the hub. We note that only $^{13}$CO (J=1-0) line is covered towards RCW~117 by ThrUMMS. We first discuss the $^{13}$CO emission followed by the results from MALT90.

\subsubsection{$^{13}$CO - ThrUMMS survey} \label{CO}
The coverage by ThrUMMS includes the hub, but do not span the filamentary structures towards the south-east of the hub. The $^{13}$CO~($J=1-0$) spectrum extracted towards the hub region is shown in Fig.~\ref{fig:spectra}. The peak emission is at $-21.3$~km~s$^{-1}$, which is consistent with the LSR velocity observed earlier \citep{2017AJ....154..140W}. We fit the observed $^{13}$CO spectrum with a Gaussian profile and find a full width at half-maximum (FWHM) of $\sim 5.0 \pm 0.1$~km~s$^{-1}$ (dispersion $\sigma\sim2.1$~km~s$^{-1}$), and peak brightness temperature of $5.5 \pm 0.1$~K. We have also constructed moment maps considering a threshold of  3$\sigma$, where $\sigma$=0.8~K. The moment-0 (intensity integrated across the profile), moment-1 (intensity weighted velocity) and moment-2 (intensity-weighted velocity dispersion) are shown in Fig.~\ref{fig:co_moments}. The integrated intensity map peaks within the hub. We also notice filamentary structures, labelled Fil-1 to Fil-6, that are visible in the column density map; shown in Fig~\ref{fig:co_moments}(a). 
From the moment-1 map, we find a variation in the centroid velocities, $\sim -19$ to $-22.5$~km~s$^{-1}$ across the region, from the southeast to the northwest (indicated by an arrow in the figure).

We notice an increase in dispersion towards the hub region from the moment-2 map (Fig.~\ref{fig:co_moments}(c)), of the order $1.8-2$~km~s$^{-1}$, as compared to a typical value of $\sim 1.0$~km~s$^{-1}$ seen for the surroundings. Considering an average dust temperature of about $30$~K towards the hub and assuming that gas and dust are coupled and have similar temperatures, the contribution to the linewidth solely due to thermal effects would be $\sim0.4$~km~s$^{-1}$. This would suggest that the observed dispersion is likely to be due to velocity gradients and/or turbulence in this region. 

\subsubsection{MALT90 survey} \label{MALT90}

Here, we present the results of the six molecular species from the MALT90 survey that are detected towards the central region of RCW~117 (displayed in Fig.~\ref{fig:malt90mom0}): HCO$^{+}$, H$^{13}$CO$^{+}$, HCN, HNC, C$_{2}$H, and N$_2$H$^{+}$. We note the presence of emission towards the central region by all species except N$_2$H$^{+}$, with the latter more prominent towards the filamentary structures in the vicinity. The spectral lines towards the peak emission, centered at $\alpha_{\textrm{J}2000}=17^h09^m35^s$, $\delta_{\textrm{J}2000}=-41^{\circ}35'52''$ within a circular region of size $26''$ are shown in Fig.~\ref{fig:spectra}. We discuss the various species below.
\\

 \noindent \textit{HCO$^{+}$ and H$^{13}$CO$^{+}$}: The HCO$^{+}$ and its isotopologue H$^{13}$CO$^{+}$ are density tracers. We find that the HCO$^{+}$ molecule has stronger red peak as compared to the blue peak. The H$^{13}$CO$^{+}$ spectrum, on the other hand, seems to demonstrate a single peak, corresponding to the LSR velocity. This would suggest that H$^{13}$CO$^{+}$ is optically thin unlike HCO$^{+}$. We also note that  HCO$^{+}$ has larger extension in velocity towards the blue side as compared to the red. We quantify the red skewed profile by estimating the asymmetry parameter $\delta$V, which is the ratio of the difference between the peak velocities of the optically thick (HCO$^{+}$) and optically thin (H$^{13}$CO$^{+}$) lines to the FWHM of the optically thin line \citep{1997ApJ...489..719M}:  
 \begin{equation}
 	\delta V=\frac{V_{thick}-V_{thin}}{\Delta V_{thin}}
 \end{equation}

 Considering $V_{thick} = -19$~km~s$^{-1}$ (peak velocity of HCO$^{+}$ line), $V_{thin} = -21.7$~km~s$^{-1}$ (central velocity of H$^{13}$CO$^{+}$ line), and $\Delta V_{thin} = 3.5$~km~s$^{-1}$, we estimate $\delta V \sim 0.8$, which is characterised as a red profile. Red asymmetry has been found towards multiple star forming regions \citep{1997ApJ...484..256G, 2008ApJ...688L..87V}, suggestive of expansion.  The integrated intensity images of HCO$^{+}$ and H$^{13}$CO$^{+}$ (Fig.~\ref{fig:malt90mom0}(a) and (d)) have an intensity peak towards C1. We also notice fainter emission towards the filamentary structures in the HCO$^{+}$ map. 
 \\

 \noindent \textit{HCN and HNC: } 
 HCN and its isomer HNC are strong tracers of dense gas. The spectral line of HCN demonstrates the red-asymmetric profile, similar to that of HCO$^{+}$. The spectrum of HNC, on the other hand, does not display any asymmetry, and peaks near the LSR velocity, suggestive of an optically thin profile. The integrated-intensity maps of HCN and HNC (Fig.~\ref{fig:malt90mom0}(b) and (e)) show enhanced emission towards C1, similar to other molecules and dust. We discern filamentary features more clearly in the HNC map as compared to the HCN. This is plausibly due to the single transition of the former as compared to the latter.
\\

  \noindent \textit{C$_{2}$H and N$_{2}$H$^{+}$: }
  C$_{2}$H is believed to be formed as a result of photodissociation of acetylene molecule, and is a very good tracer of photodissociation regions (PDRs) \citep{2012A&A...543A..27G}. However, there are evidences of formation of C$_{2}$H in denser regions via neutral reactions, and hence these molecules can act as tracers for denser gas in the region as well. In the case of C$_{2}$H, we find bright emission towards C1 in addition to localised emission to the east and west of C1 (Fig.~\ref{fig:malt90mom0}(c)). The latter correspond to emission from the ends of the bridge filament.

  Dense and cold molecular clouds are found to be traced efficiently by N$_{2}$H$^{+}$. However, the presence of CO destroys N$_{2}$H$^{+}$, forming HCO$^{+}$ \citep{2013PASA...30...57J}. The N$_{2}$H$^{+}$ spectrum is composed of 7 hyperfine components. We note a lack of emission towards C1 in the integrated intensity maps of the molecule. Rather, enhanced emission is detected towards the filamentary structures in the east and west of C1, a comparison with column density is shown in Fig.~\ref{fig:malt90mom0}(f). This is consistent with the fact that this molecule is plausibly destroyed towards C1 where HCO$^+$ emission peaks. The spectrum towards the enhanced emission, green circle shown in Fig.~\ref{fig:malt90mom0}(f) is also depicted in Fig.~\ref{fig:spectra}.
  
 \subsection{Ionized gas emission} \label{res:radio}
 
The ionized emission at 1280 and 610~MHz spans a region of size $\sim5\times3$~pc$^2$ and shows extension towards the north and east (see Fig.~\ref{fig:radiopanel}).  The total flux density is $24.2\pm2.5$~Jy and $22.5\pm2.2$~Jy for 610 and 1280~MHz, respectively, for emission integrated upto the 3$\sigma$ level ($\sigma$ is the rms noise in the image). Considering a UV range of $0.2-50$~k$\lambda$, the spectral index of the full region is found to be $-0.1$, indicative of thermal emission from the H{\small II} region. Most of the ionized gas is confined to the hub region and the peak lies close to C1. The rate of Lyman continuum photon flux ($\dot{N}_{Lyc}$) is estimated at the frequency  $\nu=1280$~MHz assuming that the emission is optically thin \citep{2016A&A...588A.143S}.

 \begin{equation} 
 \label{eqn:lyman}
 \centering
 \left( \frac{\dot{N}_{Lyc}}{s^{-1}} \right) 
 = 4.771 \times 10^{42} \left( \frac{S_{\nu}}{Jy} \right)  \left(\frac{T_{e}}{K} \right) ^ {-0.45} \left(\frac{\nu}{GHz} \right) ^ {0.1}  \left( \frac{d}{pc} \right) ^{2} 
 \end{equation}
 
 Here, $S_{\nu}$ is the integrated flux density, $d=2500$~pc is the distance to the source and $T_{e}$ is the electron temperature. We take $T_e=6960$~K from \citet{2020A&A...636A...2P} determined using free-free absorption at low radio frequencies. We estimate $\dot{N}_{Lyc}=1.26 \times 10^{49}$~s$^{-1}$ for this region which corresponds to a single main-sequence star of spectral type O6 \citep{2005A&A...436.1049M}. 
 
 In order to get an estimate of the variation of electron density across the large region, we have generated the electron density map of the region using the 1280~MHz map, after convolving it with a beam of size $13''$ and regridding to a pixel size of $4''$. For this, we assume optically thin emission and negligible absorption by dust. We considered the pixels with values greater than $3\sigma$ to generate the map. The value of electron density $n_{e}$ has been calculated for each pixel using the following expression \citep{rybicki.lightman.ch5}: 
 \begin{equation}
 	\centering
 	n_{e} \textrm{ (cm\textsuperscript{-3})} = \left( \frac{1.47 \times 10^{37} 4 \pi d^{2} S_{\nu}e^{\frac{h\nu}{k_{B}T_{e}}}T_{e}^{1/2}}{dV\overline{g}_{ff}} \right)^{1/2}
 \end{equation}
 \noindent Here, $dV$ is the volume over which electron density is determined, $k_B$ is the Boltzmann's constant and $h$ is the Planck's constant. $\overline{g}_{ff}$ is the velocity averaged Gaunt factor, taken to be 5 as $u = h\nu/k_B T_e = 8 \times 10^{-6}$ \citep{rybicki.lightman.ch5}. The volume $dV$ is estimated by multiplying the pixel area with the line of sight distance which we take to be uniform ($6.5$~pc). This represents the approximate size of the region. The largest value of $n_{e}$, which is obtained towards the radio peak is $\sim750$~cm$^{-3}$, while the average value across the region comes out to be $\sim300$~cm$^{-3}$.
 
 \subsection{Young Stellar Objects in mid-infrared } \label{sec:result_yso}

   \begin{figure}
 	\centering
 	\includegraphics[width=0.95\columnwidth]{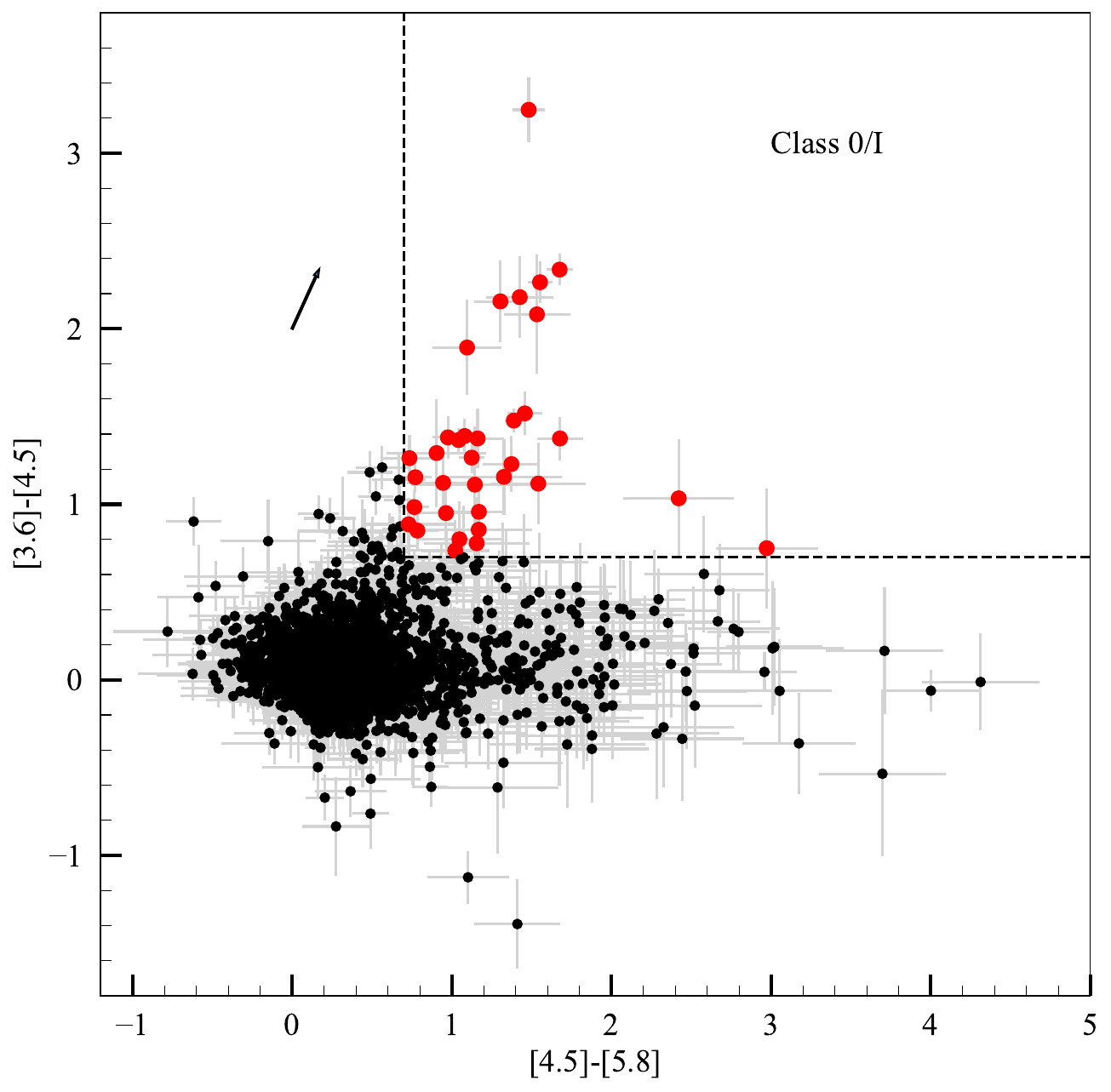}
 	\caption{The \textit{Spitzer} colour-colour diagram for sources extracted from a region spanning $18 \arcmin \times 18 \arcmin$ towards RCW~117. The arrow indicates the reddening vector for $A_{V}$=20~mag, using the extinction law of \citet{2007ApJ...663.1069F}. The Class 0/I sources (red circles) towards the top right have been demarcated using dotted black lines.}
  \label{fig:YSO_spitzer}
   \end{figure} 
 

 We have extracted point sources towards RCW~117 from the Spring '01 GLIMPSE 07 Archive \citep{2003PASP..115..953B}. We use the mid-infrared colours [3.6 - 4.5] versus [4.5 - 5.8] to identify the young stellar objects (YSOs) in the region. These bands are particularly sensitive to emission from circumstellar envelopes and disks \citep{2004ApJS..154..363A, 2004ApJS..154..367M}. We utilise only the three bands: 3.6, 4.5 and 5.8~$\mu$m, as the longer 8~$\mu$m image suffers from saturation towards pixels in the hub region where the emission peaks. In addition, the presence of large scale diffuse emission makes the source extraction difficult. Here, we focus only on the youngest Class 0/I YSOs as we would like to compare their locations relative to the dense molecular gas cores and filaments. Class 0/I YSOs are deeply embedded sources associated with a central protostar and disk surrounded by an envelope. These sources are selected using the colors [3.6 - 4.5] versus [4.5 - 5.8] based on \citet{2005ApJ...629..881H, 2012ApJ...755...20S}.
We consider a region of size $18'\times18'$ that encompasses the filamentary features associated with RCW~117, shown in Fig.~\ref{fig:YSO_spitzer} and find 3486 objects that are detected in the three selected bands. Amongst these, we find that only 34 objects can be categorised as Class 0/I objects. These are tabulated in Appendix A, Table.~\ref{tab:classi_yso_table}. We note that most of the GLIMPSE sources are located away from the hub where the background emission is low. While we find 9 YSOs within the hub, we note a few compact sources lying towards the centre within the nebulous emission varying across the region. However, these sources are not present in the catalog, plausibly due to the lack of accurate photometry in the presence of nebulous emission.

   \begin{figure*}
 	\centering
 	\includegraphics[width=\textwidth]{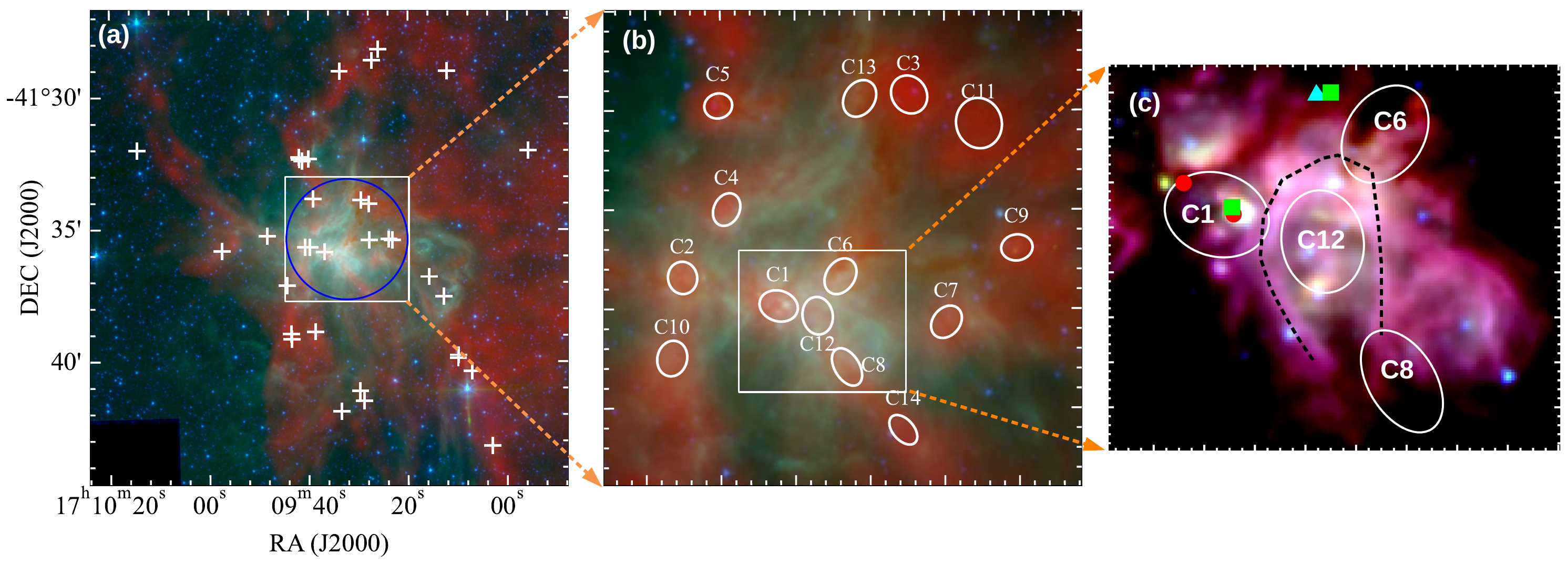}
 	\caption{(a) Three-colour composite image of RCW~117, with the column density image in red, the \textit{Spitzer} $5.8~\mu$m image in green and the \textit{Spitzer} $3.6~\mu$m in blue. The detected Class 0/I YSOs are represented as white crosses. The blue circle marks the hub. (b) Enlarged view of central region with the same three-colour composite as (a). The identified cores are marked as ellipses with the size same as the deconvolved core size. (c) \textit{Spitzer} three-colour composite image with $5.8~\mu$m as red, $4.5~\mu$m as green, and $3.6~\mu$m as blue of the rectangular region depicted in (b) is shown. The hydroxyl masers are depicted as green squares, the H$_{2}$O maser as a cyan triangle, and the methanol masers using red circles. The horseshoe shaped structure enveloping C12 is indicated by the black dashed curve.}
  \label{fig:multiband_rgb}
   \end{figure*} 
 


\begin{figure}
	\includegraphics[width=\columnwidth]{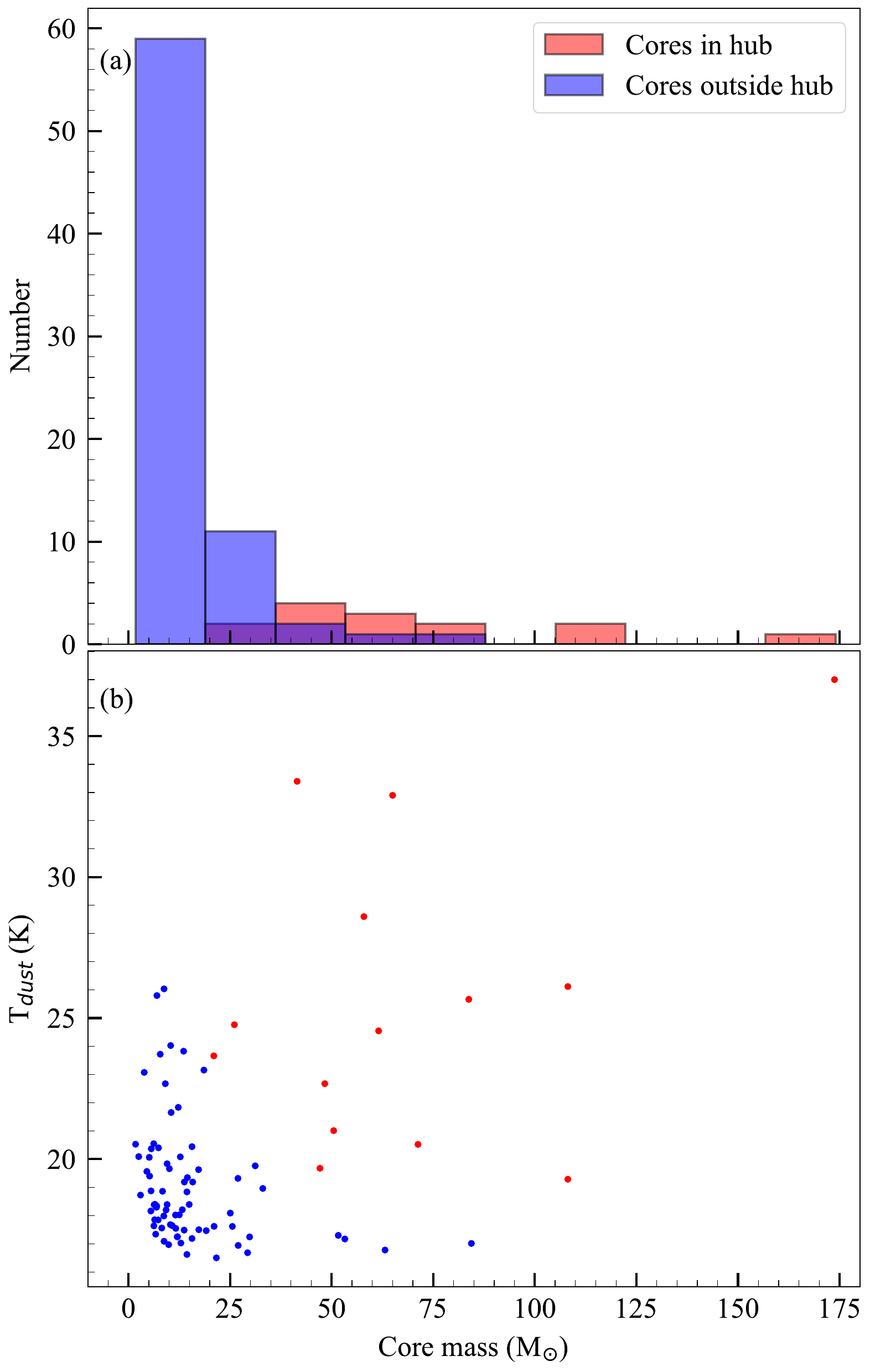}
	\caption{(a) Distribution of cores located within (red) and outside (blue) the hub. (b) Plot of the mean core temperature versus the core mass. The cores within the hub are marked as red filled circles, while the rest of the cores are marked as blue filled circles.}
	\label{fig:histo}
\end{figure} 


  \section{Discussion} \label{sec:discussions}

The hub-filament system associated with RCW~117 has been studied comprehensively in the infrared, and radio continuum wavebands. It has been suggested that hubs have the potential to form star clusters, with the likely presence of massive stars \citep{2009ApJ...700.1609M, 2019A&A...629A..81T, 2022A&A...658A.114K}. In our case, there is a large possibility of massive stars being present near the hub, which is reinforced by the fact that the ionised gas emission in radio wavebands peaks towards this region and massive cores are found here.


\begin{figure}
	\includegraphics[width=\columnwidth]{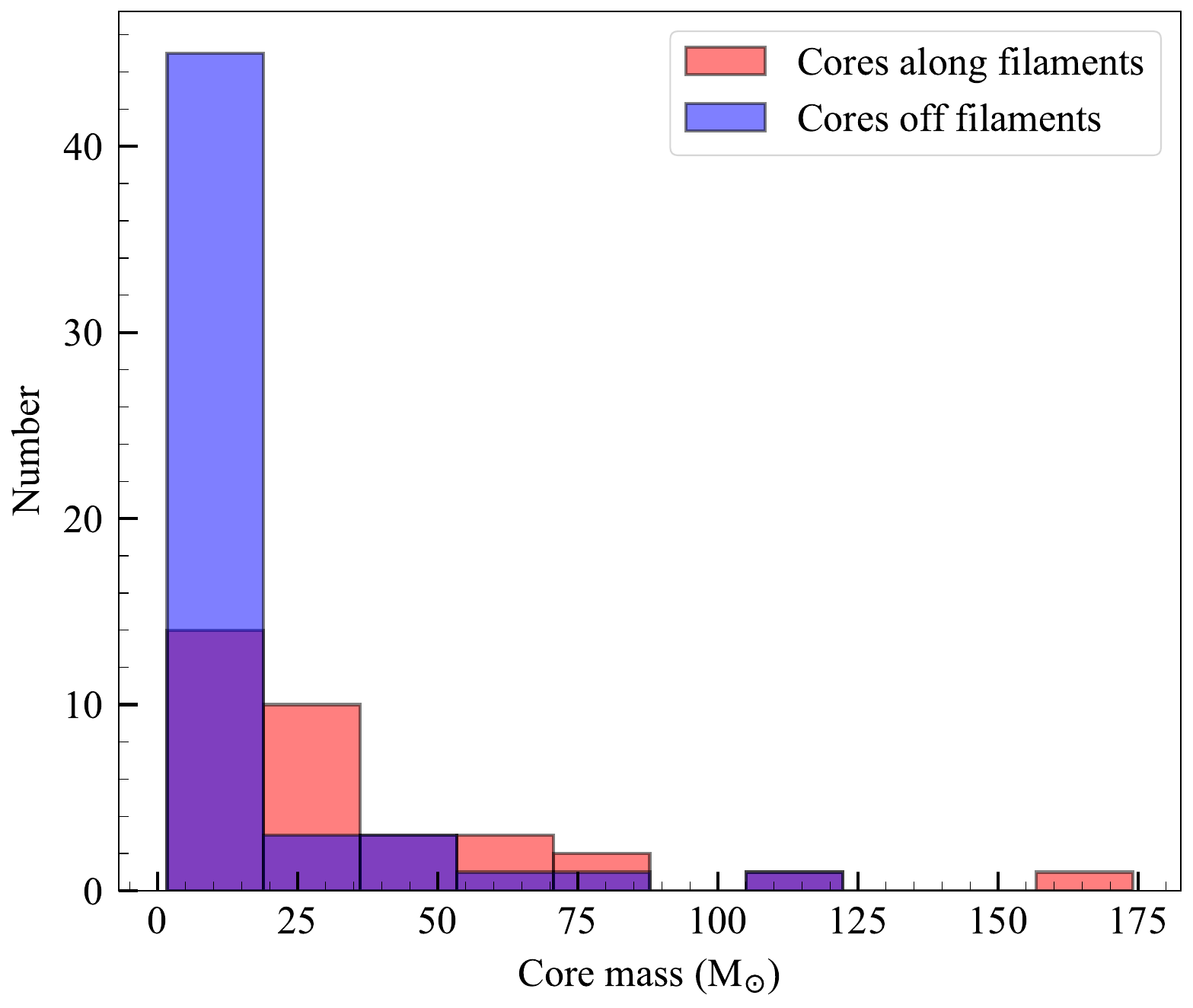}
	\caption{Histogram of cores located along the filaments (red) as compared to those located away from the filaments (blue).}
	\label{fig:frag_histo}
\end{figure} 


   \begin{figure*}
 	\centering
    \includegraphics[width=\textwidth]{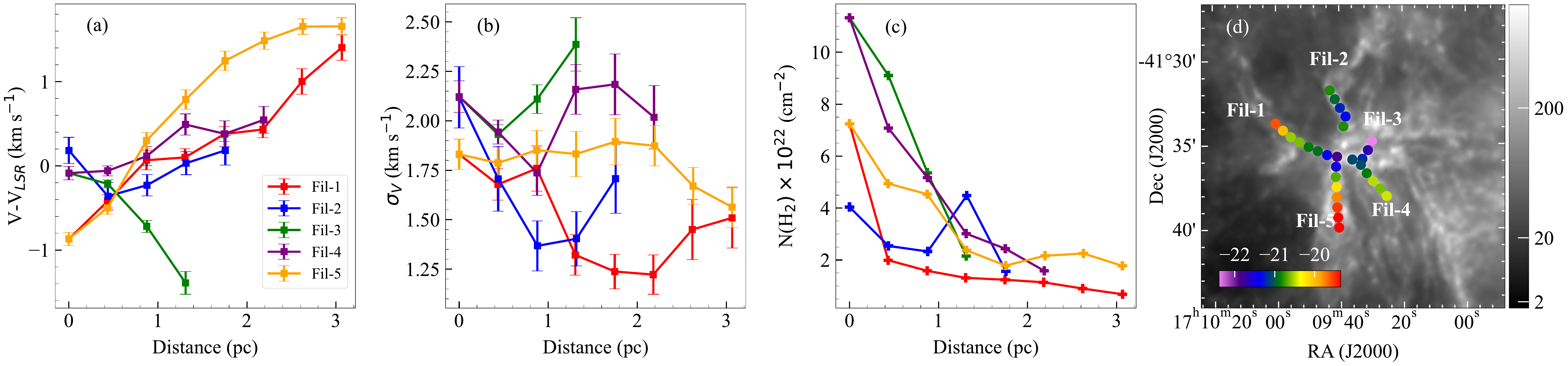}
    \caption{(a) Variation of velocity (relative to the LSR velocity), calculated using the $^{13}$CO ($J=1-0$) line, along the filament for Fil-1 to Fil-5, which represent filaments converging to the hub. The origin corresponds to the end of the filament located nearest to the hub. (b) Variation in the velocity dispersion along the filaments. (c) Variation of mean column densities along the filaments. (d) Greyscale image is the column density in units of $10^{20}$~cm$^{-2}$, with the velocities of filaments marked as coloured circles, in units of km~s$^{-1}$. }
    \label{fig:velocity_gradients}
   \end{figure*} 

\subsection{Star formation within the hub} \label{disc:hub}
In this section, we analyse the star-formation activity within the hub. As a first step, we compare the mass of the 14 cores present within the hub with masses of those present in the outer regions, most of which are located within the filaments. This comparison is shown in Fig.~\ref{fig:histo}(a). We find that the cores within the hub lie on the higher end of the mass range with masses between $21 - 174$~M$_{\odot}$. On the other hand, the 74 cores lying outside the hub region have masses in the range, $2-84$~M$_{\odot}$ with nearly 80\% of these cores having masses $<20$~M$_{\odot}$. We, therefore, infer that the cores in the hub are more massive than those lying in the outer regions.

The cores located within the hub are also hotter, $20-37$~K as compared to the cores located outside, $16-26$~K. The distribution of the mean core temperatures is displayed in Fig.~\ref{fig:histo}(b). This would suggest that the cores in the hub are more evolved with higher star-formation activity, in general, as compared to the other cores located outside. We also compare the distribution and association of Class 0/I sources within the hub. Of the 34 Class 0/I sources across the full region, 9 sources (26\%) are located within the hub, associated with cores C1, C2, C3, C5, and C9. The presence of these Class 0/I sources towards the cores signifies that the corresponding cores are in an active star-forming state.

As mentioned earlier in Sec.~\ref{sec:warmdust}, the warm dust emission towards the hub is quite bright and shows numerous filamentary and arc-like structures. As seen in Fig.\ref{fig:multiband_rgb}(a), the distribution of ionised gas emission from the region is similar to the morphology of the warm dust emission. In the vicinity of the core C12, we find a horseshoe shaped arc-like structure encompassing it towards the north, visualised in Fig.\ref{fig:multiband_rgb}(c). It is plausible that this arc is due to shocked gas from stellar winds in the interior leading to the formation of a bubble-like structure. 

The core C1, represents the location of the brightest MIR emission in \textit{Spitzer} bands that has a point-like emission. This point source is listed in the catalog in only two bands of 4.5 and 5.8~$\mu$m bands with magnitude 4.76 and 4.70 mags, respectively.
A Class II methanol maser \citep{1995MNRAS.272...96C} and a hydroxyl maser \citep{2003MNRAS.341..551C} have been identified towards the location of MIR peak. Class II methanol masers are often located towards hot molecular cores, UC H{\small II} regions, and OH (hydroxyl) masers which, in turn, are found towards massive star forming regions \citep{1991ASPC...16..119M}. The presence of the methanol maser towards C1 suggests that the core is in an active star forming state. We also find two hydroxyl masers \citep{1995MNRAS.273..328C, 1995MNRAS.274..808C} and a water maser \citep{1983AuJPh..36..401C} towards the central hub (see Fig.~\ref{fig:multiband_rgb}). 
The ionised gas emission at 1280~MHz peaks 
in the vicinity of the C1 and C12. 
However, as the radio emission could also suffer from local optical depth effects, it is difficult to pin-point the exact location of exciting stars. Considering that the radio and MIR emission peak towards central region of the hub, and hosts masers, it is reasonable to conclude that massive stars are embedded here.
   
\subsection{Global star formation scenario within RCW~117} \label{disc:scenario}

It is known that molecular clouds begin their journey from thin cold atomic gas sheets and evolve to clumps and cores, eventually forming stars \citep{2018ARA&A..56...41M, 2019MNRAS.485.4686W}. The parent clouds are ultimately dispersed by massive stars that are formed within \citep{1979MNRAS.186...59W, 2008ASPC..387..148M}. In the case of RCW~117, the \textit{Herschel} observations have revealed a  hub-filamentary structure, with six prominent filaments (Fil-1 to Fil-6) within the hub. Twenty one of the 88 cores are located along the prominent filaments Fil-1 to Fil-6. Overall, we find that $\sim 39$~percent of the detected cores are located along the filaments (see Fig.~\ref{fig:frag_histo}). This strongly suggests accretion and fragmentation along the lengths of the filaments \citep{2010A&A...518L.102A, 2010A&A...518L.106K}. It is to be noted that as the mass of the lowest core detected is 1.7~M$_\odot$, it is very likely that the core population sampled in this study is incomplete as this does not cover the entire range of core masses. For instance, the minimum mass of core(s) identified by \citet{2015A&A...584A..91K} towards Aquila is 0.01~M$_\odot$ while \citet{2013ApJ...777L..33P} estimate the median value towards dense cores and filaments in Orion-A to be $\sim 0.2$~M$_\odot$. Similarly, the filament fragments observed due to visual inspection may be constituents of weaker filaments that are not detected. 

We estimate the velocities along the length of the filaments Fil-1 to Fil-5, by analysing the $^{13}$CO ($J=1-0$) spectra integrated within circular regions of size $36$~arcsec along the filament. The velocity structures across the filaments are shown in Fig.~\ref{fig:velocity_gradients}. In the Fig~\ref{fig:velocity_gradients}(a), (b) and (c), the X-axis represents the distance from the end of the filamentary structure closest to the hub (proximal end). We observe that the velocities towards the proximal end are blue-shifted as compared to the ends located away from the hub (distal end) for filaments other than Fil-3, see Fig~\ref{fig:velocity_gradients}(a).
For the latter, we note that it is a short filament which converges towards core C6 in the hub. It is not clear why this filament shows a differing velocity gradient, and higher resolution images would aid in clarifying this. We also plot the variation in the velocity dispersion along the  filament in Fig.~\ref{fig:velocity_gradients}(c). We find that the filaments display larger dispersion towards the proximal end, as compared to the distal end, with the exception being Fil-3. Higher dispersion towards the hub can be attributed to gravitational acceleration \citep{2020ApJ...905..158W}, or to outflow-generated turbulence, which results in a local increase in the velocity dispersion \citep{2012A&A...543A.140D}.

 Column densities along all the  filaments are observed to decrease from the proximal to the distal end (see Fig.~\ref{fig:velocity_gradients}(c)), which is in tandem with the observed velocity gradients, and provides strong evidence for filamentary accretion \citep{2013ApJ...766..115K, 2014A&A...565A.101T}. We estimate the mass accretion rate ($\dot{M}_{||}$) along the filaments using the following expression \citep{2013ApJ...766..115K}:
\begin{equation}
    \dot{M}_{||}=\frac{\nabla V_{||, obs}M}{tan(\alpha_{inc})}
\end{equation}
\noindent Here, $\nabla V_{||, obs}$ is the velocity gradient observed along the filament, $M$ is the mass of the filament, and $\alpha_{inc}$ is the inclination angle of the filament with respect to the plane of sky. $\nabla V_{||, obs}$ is estimated by means of difference in velocity at the edges of the filament ($\sim0.3-1$~km~s$^{-1}$~pc$^{-1}$), and we take $\alpha_{inc}\sim 45^{\circ}$ as a nominal value for the inclination angle. The velocity gradients and mass-accretion rate estimates are listed Table~\ref{tab:velograd}. The accretion rates lie in the range $1.6 \times 10^{-4}-10^{-3}$~M$_{\odot}$~yr$^{-1}$, with a mean value of $\sim 5.36 \times 10^{-4}$~M$_{\odot}$~yr$^{-1}$. These values can change by a factor $0.58$ to $1.73$ for filament inclinations varying between $30^{\circ}$ and $60^{\circ}$, respectively. We exclude Fil-3 for estimation of mass-accretion rate as it does not show an increasing velocity gradient towards the proximal end. We also perceive a velocity gradient ($\sim3$~km~s$^{-1}$) on larger spatial scales in the cloud along the northeast-southwest direction in the $^{13}$CO moment-1 map (see Sect.~\ref{CO}). This can plausibly arise due to a number of physical processes such as collapse, expansion, rotation about an axis passing through the hub or cloud-cloud collision \citep{2012ApJ...748...16T, 2012ApJ...746..174R, 2015A&A...580A..49I, 2018ApJ...859..166F}.


\begin{table}
	\centering
	\caption{Properties of filaments converging to the hub. Here V$_{||, obs}$ is the observed velocity gradient, M$_{fil}$ is the filament mass, and $\dot{M}_{||}$ is the accretion rate}
	\begin{tabular}{cccc} 
		\hline \hline
		Fil. ID & M$_{fil}$~(M$_{\odot}$) & $\nabla$V$_{||, obs}$ (km~s$^{-1}$~pc$^{-1}$) & $\dot{M}_{||}$ ($\times 10^{-4}$~M$_{\odot}$ yr$^{-1}$) \\
		\hline
		Fil-1 & 547.3 & 0.74 & 4.14 \\
        Fil-2 & 375.6 & 0.41 & 1.58 \\
        Fil-3 & 946.5 & -0.99 & - \\
        Fil-4 & 1041.7 & 0.29 & 3.08 \\
        Fil-5 & 855.3 & 0.95 & 8.39 \\
	\hline \hline
	\end{tabular}
 \label{tab:velograd}
\end{table}


 A comparison of the distribution of cores located along the filaments with the others that appear to be isolated, indicates that 34 out of 88 cores are located along the filaments. These cores have a wide mass distribution (8-108~M$_\odot$), as depicted in Fig.~\ref{fig:frag_histo}. In addition, we find many Class 0/I YSOs lying towards these cores. The formation and growth of cores in accreting filaments has been explored by a number of simulations and theoretical studies, implying that stars can be formed in cores formed through the process of filament fragmentation \citep{2021MNRAS.502..564A, 2019A&A...628A.112K}. 
It is, therefore, likely that the cores located on the filaments in RCW~117 are also produced due to internal perturbations  leading to their fragmentation \citep{1964ApJ...140.1529O}.
 Studies of star-formation in other nearby filamentary molecular clouds in the \textit{Herschel} Gould Belt Survey also find cores located along the filaments, thus supporting this scenario of filament fragmentation \citep{2010A&A...518L.102A, 2015A&A...584A..91K}. A lower age limit can be ascertained for fragmenting filaments where cores are located at regular spatial intervals along the filament based on the assertion of gravo-acoustic oscillations along the filament, as follows \citep{2016MNRAS.458..319C}:

    \begin{equation}
    \tau_{age} > \tau_{crit} \simeq \frac{\lambda_{core}}{2a_{0}}
    \end{equation}

\begin{table}
	\centering
	\caption{Properties of filaments with cores along their length suggesting fragmentation. $\lambda_{core,median}$ represents the spacing between cores along the filament, $a_0$ is the sound speed associated with the filament and $\tau_{crit}$ is the lower limit to the age of the filament, see text for more details.}
	\begin{tabular}{cccc} 
		\hline \hline
		Fil. ID & $\lambda_{core,median}$ (pc)& $a_0$ (km~s$^{-1}$) & $\tau_{crit}$ (Myr)\\
		\hline
		Fil-1 & 0.54 & 0.26 & 1.02\\
        Fil-2 & 0.51 & 0.23 & 1.08\\
        Fil-3 & 0.33 & 0.28 & 0.57\\
        Fil-4 & 0.59 & 0.26 & 1.1\\
        Fil-5 & 0.39 & 0.26 & 0.73\\
        Fil-6 & 0.51 & 0.26 & 0.95\\
        Fil-7 & 0.48 & 0.24 & 0.97\\
        Fil-8 & 0.45 & 0.22 & 1.00\\
        Fil-9 & 0.49 & 0.23 & 1.04\\
        Fil-10 & 1.29 & 0.23 & 2.74\\
        Fil-11 & 0.58 & 0.22 & 1.28\\
	\hline \hline
	\end{tabular}
 \label{tab:fragmentation}
\end{table}


  \noindent Here $\lambda_{core}$ represents the spacing between cores along a certain filament, and $a_{0}$ is the isothermal sound speed within that filament. We take the median value of separation between cores along each filament for the analysis as we do not find that the cores are at uniform separation from each other along most filaments. Amongst the filaments, we consider Fil-1 to Fil-11 for this analysis as Fil-12 does not show the presence of cores along its length. The values of core separation, sound speed and lower limit to the age of each filament is listed in Table~\ref{tab:fragmentation}. We find $\lambda_{core}\sim 0.4 - 0.6$~pc for filaments other than Fil-10 which has $\lambda_{core}\sim 1.3$~pc. We note that there are only two cores along this filament (Fil-10) and they are located at the edges of the filament strongly pointing towards the edge-dominated collapse scenario proposed for filamentary structures \citep{2020A&A...637A..67Y, 2022MNRAS.517.5272H}. The isothermal sound speed is about $a_{0}\sim0.2-0.3$~km~s$^{-1}$. With these values, we assess the age of the filaments to be $\sim1$~Myr or larger, thus providing us with a lower limit to the age of the cloud. 
   
  When we attempt to visualize the overall picture of RCW~117 based on various tracers considered in this study, we find a number of salient features: (i) a large network of filamentary structures, (ii) accreting filaments 
converging to a hub, (iii) the hub is the location where massive stars are forming, (iv) filaments fragmenting to cores, and (v) isolated cores that do not appear to be associated with filaments. All these features can plausibly be explained with the help of the Global Hierarchical Collapse (GHC) scenario \citep{2017MNRAS.467.1313V, 2019MNRAS.490.3061V}. According to this model,  the collapse is initiated by compressions in the warm atomic phase, which grow and transition to the cold atomic phase. These compressions which make up the cloud, accrete more material from the surroundings, thus lowering the Jeans' mass. This results in pressureless collapse along the shortest dimension in the cloud, resulting in the formation of filaments. Longitudinal flow along these filaments creates cores, which increase in mass and size, causing more matter to accrete onto the filaments. As the collapse proceeds, the rate and number of collapsing objects increases. This increases the star formation rate, leading to the formation of massive stars towards the hub. According to GHC, the non-thermal motions observed in clouds are likely to consist of a combination of infall and truly turbulent motions, resulting in chaotic motion. Further, the formation of newly formed massive stars can lead to cloud disruption locally towards the hub, leading to the dispersal of filaments over time. This is consistent with the presence of expansion signature towards the centre of the hub.

 \begin{figure}
 	\includegraphics[width=\columnwidth]{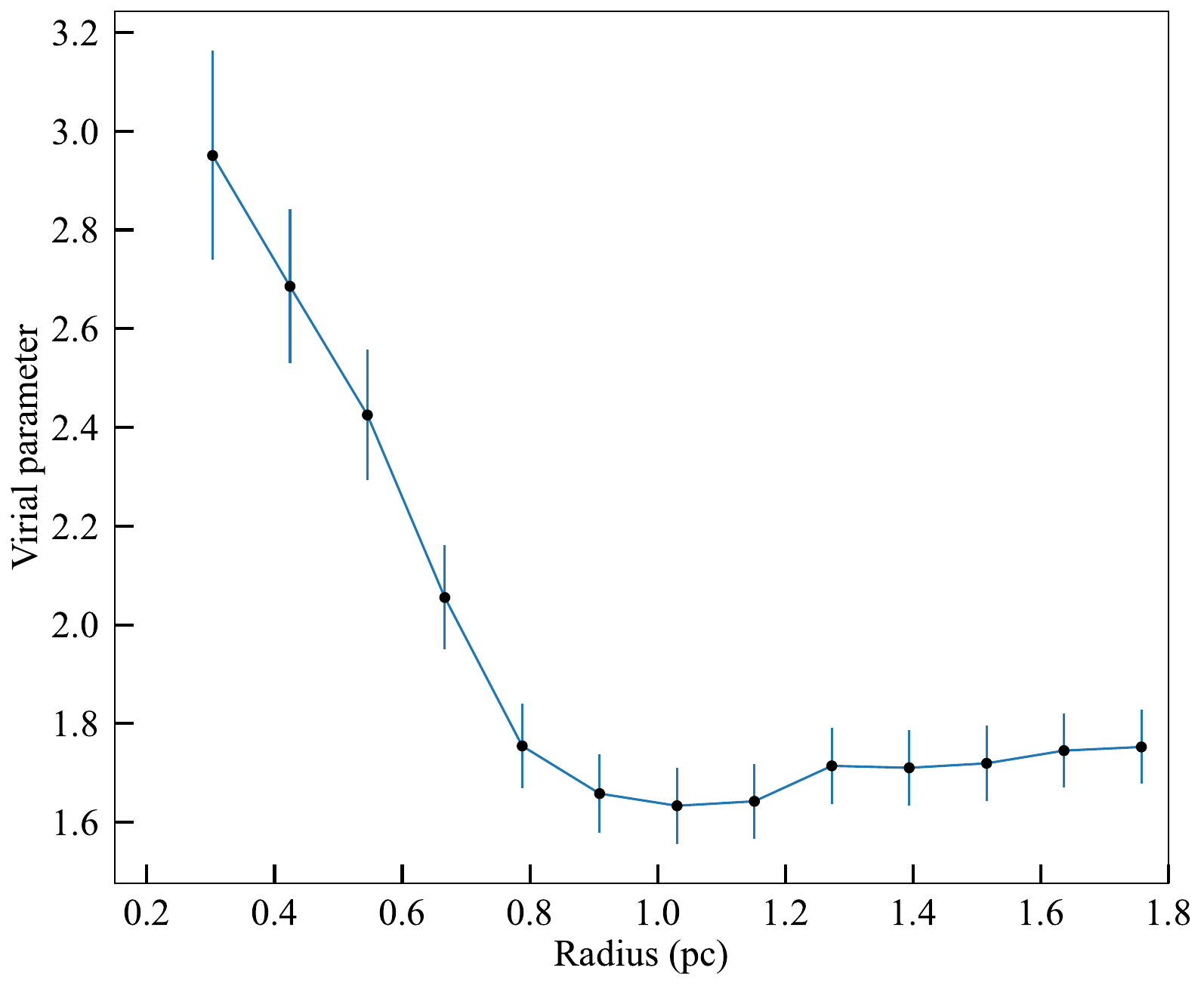}
 	\caption{The virial parameter profile estimated using $^{13}$CO ($J=1-0$) line for the central region using the concentric spheres centered at the hub.}
 \label{fig:virial_param}
 \end{figure} 


Towards RCW~117, we believe that GHC can help explain the star-formation scenario. The multiple cores located along the filaments suggest that filament fragmentation is occurring in this region. The velocity gradients observed along filaments converging to the hub suggest a longitudinal flow of matter, feeding the massive star formation in the hub as well as the core growth along the filaments. Towards the hub, we find a marginal increase in the velocity dispersion, suggesting that the region is more turbulent as compared to its surroundings. This increased turbulence may be due to the feedback generated by the massive stars located inside cores within the hub which corroborates with the lack of infall signature observed here. 

To confirm this, we assess the virial parameter ($\alpha_{vir}$) for the central hub region vis-a-vis the cloud using the $^{13}$CO emission. The virial parameter is estimated using $\alpha_{vir}=\frac{5 \sigma_{v}^{2}R}{GM}$, where $\sigma_{v}$ is the velocity dispersion, $R$ is the radius of the region under consideration, and $M$ is the corresponding mass. We construct a virial parameter profile by considering circular regions concentric about the core C1 in the hub ($\alpha_{\textrm{J}2000}=17^{h}09^{m}35.75^{s}$, $\delta_{\textrm{J}2000}=-41^{\circ}35'57.17''$). The circular regions have increasing radius from $25''$ to $145''$, in steps of $10''$. The $\alpha_{vir}$ profile is shown in Fig.~\ref{fig:virial_param}. We find that the $\alpha_{vir}$ is highest towards the central region in the hub ($\sim 3$), and decreases outwards upto 1~pc, after which it remains more or less constant, at $\sim1.8$. This implies that the innermost hub region may be affected by feedback effects from the massive stars and protostars within, resulting in an increase in the virial parameter here, while the outer regions suggest infall when the total emission is considered. This would indicate scale dependant collapse, where the outer cloud is collapsing while this signature is not seen towards the centre due to feedback effects. An alternate possibility is that this profile could be the outcome of projection effects of non-homogeneous radial density distribution of hub and filaments, as similar virial ratio profiles are obtained for a sample of IRDCs by \citet{2023MNRAS.525.2935P}.

Finally, we note that the cores that appear to be isolated are probably remnants of filament fragmentation where the filaments have dispersed or are in the process of dispersal with emission too weak to be detected. The hub represents the region of massive star-formation flurry with ionised gas emission, presence of masers and multiple mid-infrared arcs indicating shocked structures. Higher spatial resolution observations should help locate the embedded cores and ascertain their evolutionary stages in more detail.

   \section{Conclusions} \label{sec:conclusions}
   We have carried out a multiwavelength study of the H{\small II} region RCW~117. Based on the investigation, we arrive at the following conclusions: 
   
   (1) The \textit{Herschel} maps reveal  extensive filamentary structures around a central hub. Column density and dust temperature maps have been generated using the \textit{Herschel} images, with higher column densities ($\sim 10^{23}$~cm$^{-2}$) observed towards the hub as compared to the filaments ($\sim 10^{22}$~cm$^{-2}$). The dust temperature ranges between $20-40$~K, with the hub having higher temperatures. 
   
   (2) Eighty eight cores and 12 filaments have been identified using the column density map. Of these, 14 cores lie within the hub, 34 cores are observed along filaments, while the rest appear isolated. The core C1, which lies towards the centre of the hub is the most massive ($174$~M$_{\odot}$) and hottest (37~K). Five filaments (Fil-1 to Fil-5) are connected to the hub, while Fil-6 appears as a bridge filament hosting the core C1. 
   
   (3) Ionised gas emission towards the region is mapped using GMRT at 1280 and 610~MHz, with total flux densities of $\sim 24.2$ and $\sim 22.5$~Jy, respectively. The electron density map generated using the 1280~MHz map has a peak density of 750~cm$^{-3}$ towards the hub, with an average value of 300~cm$^{-3}$ across the region. 
   
   (4) The warm dust emission, mapped by \textit{Spitzer} MIR wavebands reveals multiple regions of shocked emission towards the hub, confirming active star formation within. The \textit{Spitzer} GLIMPSE archive has been used for identifying YSOs towards the region. Thirty-four Class 0/I YSOs are identified with many of them lying within the hub and along the filamentary structures.
   
   (5) The \textit{ThrUMMS} $^{13}$CO integrated intensity map shows similar morphological features i.e the hub and surrounding filamentary features (Fil-1 to Fil-6 in particular) as the \textit{Herschel} images. The moment-1 map shows a velocity variation along the southwest-northeast direction of $\sim 3$~km~s$^{-1}$, while the moment-2 map shows enhanced velocity dispersion towards the hub as compared to the surroundings. Velocity gradients in the range $\sim 0.3 - 1$~km~s$^{-1}$~pc$^{-1}$ are observed towards filaments directly connected to the hub (Fil-1 to Fil-5), suggesting longitudinal flow of matter towards the hub. 
   
   (6) Based on the distribution of cores and YSOs, and the velocity gradients observed along the filaments, we propose the global hierarchical collapse (GHC) model as a plausible star formation scenario within RCW~117. The model proposes the flow of matter along filaments onto the central hub, and also the fragmentation of filaments due to longitudinal accretion of matter. It also explains the star formation markers observed towards the hub such as masers, shocked emission, and ionised gas emission. 

\section*{Acknowledgements}
We thank the referee for providing suggestions that have improved the quality of the paper. We thank the staff of the GMRT that made these observations possible. GMRT is run by the National Centre for Radio Astrophysics of the Tata Institute of Fundamental Research. This publication makes use of data from Herschel, which is an ESA space observatory with science instruments provided by European-led Principal Investigator consortia and with important participation from NASA. This work is based [in part] on observations made with the Spitzer Space Telescope, which was operated by the Jet Propulsion Laboratory, California Institute of Technology under a contract with NASA. We acknowledge the support received from Alexander Men'shchikov of the Université Paris-Saclay for clarifications regarding the software package \textit{getsf}. Arun Seshadri and Sarita Vig acknowledge financial support from the Department of Science and Technology - Science and Engineering Research Board (SERB) grant CRG/2019/002581. D.K. Ojha acknowledges the support of the Department of Atomic Energy, Government of India, under Project Identification No. RTI 4002.

\section*{Data Availability}
The data underlying this article will be shared on reasonable request to the corresponding author.



\bibliographystyle{mnras}
\bibliography{example} 

\begin{thebibliography}{}
\makeatletter
\relax
\def\mn@urlcharsother{\let\do\@makeother \do\$\do\&\do\#\do\^\do\_\do\%\do\~}
\def\mn@doi{\begingroup\mn@urlcharsother \@ifnextchar [ {\mn@doi@} {\mn@doi@[]}}
\def\mn@doi@[#1]#2{\def\@tempa{#1}\ifx\@tempa\@empty \href {http://dx.doi.org/#2} {doi:#2}\else \href {http://dx.doi.org/#2} {#1}\fi \endgroup}
\def\mn@eprint#1#2{\mn@eprint@#1:#2::\@nil}
\def\mn@eprint@arXiv#1{\href {http://arxiv.org/abs/#1} {{\tt arXiv:#1}}}
\def\mn@eprint@dblp#1{\href {http://dblp.uni-trier.de/rec/bibtex/#1.xml} {dblp:#1}}
\def\mn@eprint@#1:#2:#3:#4\@nil{\def\@tempa {#1}\def\@tempb {#2}\def\@tempc {#3}\ifx \@tempc \@empty \let \@tempc \@tempb \let \@tempb \@tempa \fi \ifx \@tempb \@empty \def\@tempb {arXiv}\fi \@ifundefined {mn@eprint@\@tempb}{\@tempb:\@tempc}{\expandafter \expandafter \csname mn@eprint@\@tempb\endcsname \expandafter{\@tempc}}}

\bibitem[\protect\citeauthoryear{{Allen} et~al.,}{{Allen} et~al.}{2004}]{2004ApJS..154..363A}
{Allen} L.~E.,  et~al., 2004, \mn@doi [\apjs] {10.1086/422715}, \href {https://ui.adsabs.harvard.edu/abs/2004ApJS..154..363A} {154, 363}

\bibitem[\protect\citeauthoryear{{Anathpindika} \& {Francesco}}{{Anathpindika} \& {Francesco}}{2021}]{2021MNRAS.502..564A}
{Anathpindika} S.~V.,  {Francesco} J.~D.,  2021, \mn@doi [\mnras] {10.1093/mnras/staa4007}, \href {https://ui.adsabs.harvard.edu/abs/2021MNRAS.502..564A} {502, 564}

\bibitem[\protect\citeauthoryear{{Anderson} et~al.,}{{Anderson} et~al.}{2021}]{2021MNRAS.508.2964A}
{Anderson} M.,  et~al., 2021, \mn@doi [\mnras] {10.1093/mnras/stab2674}, \href {https://ui.adsabs.harvard.edu/abs/2021MNRAS.508.2964A} {508, 2964}

\bibitem[\protect\citeauthoryear{{Andr{\'e}} et~al.,}{{Andr{\'e}} et~al.}{2010}]{2010A&A...518L.102A}
{Andr{\'e}} P.,  et~al., 2010, \mn@doi [\aap] {10.1051/0004-6361/201014666}, \href {https://ui.adsabs.harvard.edu/abs/2010A&A...518L.102A} {518, L102}

\bibitem[\protect\citeauthoryear{{Andr{\'e}}, {Di Francesco}, {Ward-Thompson}, {Inutsuka}, {Pudritz}  \& {Pineda}}{{Andr{\'e}} et~al.}{2014}]{2014prpl.conf...27A}
{Andr{\'e}} P.,  {Di Francesco} J.,  {Ward-Thompson} D.,  {Inutsuka} S.~I.,  {Pudritz} R.~E.,   {Pineda} J.~E.,  2014, in {Beuther} H.,  {Klessen} R.~S.,  {Dullemond} C.~P.,   {Henning} T.,  eds, Protostars and Planets VI. p.~27 (\mn@eprint {arXiv} {1312.6232}), \mn@doi{10.2458/azu\_uapress\_9780816531240-ch002}

\bibitem[\protect\citeauthoryear{{Andr{\'e}}, {Palmeirim}  \& {Arzoumanian}}{{Andr{\'e}} et~al.}{2022}]{2022A&A...667L...1A}
{Andr{\'e}} P.~J.,  {Palmeirim} P.,   {Arzoumanian} D.,  2022, \mn@doi [\aap] {10.1051/0004-6361/202244541}, \href {https://ui.adsabs.harvard.edu/abs/2022A&A...667L...1A} {667, L1}

\bibitem[\protect\citeauthoryear{{Andrews}, {Boersma}, {Werner}, {Livingston}, {Allamandola}  \& {Tielens}}{{Andrews} et~al.}{2015}]{2015ApJ...807...99A}
{Andrews} H.,  {Boersma} C.,  {Werner} M.~W.,  {Livingston} J.,  {Allamandola} L.~J.,   {Tielens} A.~G.~G.~M.,  2015, \mn@doi [\apj] {10.1088/0004-637X/807/1/99}, \href {https://ui.adsabs.harvard.edu/abs/2015ApJ...807...99A} {807, 99}

\bibitem[\protect\citeauthoryear{{Arzoumanian} et~al.,}{{Arzoumanian} et~al.}{2011}]{2011A&A...529L...6A}
{Arzoumanian} D.,  et~al., 2011, \mn@doi [\aap] {10.1051/0004-6361/201116596}, \href {https://ui.adsabs.harvard.edu/abs/2011A&A...529L...6A} {529, L6}

\bibitem[\protect\citeauthoryear{{Arzoumanian}, {Andr{\'e}}, {Peretto}  \& {K{\"o}nyves}}{{Arzoumanian} et~al.}{2013}]{2013A&A...553A.119A}
{Arzoumanian} D.,  {Andr{\'e}} P.,  {Peretto} N.,   {K{\"o}nyves} V.,  2013, \mn@doi [\aap] {10.1051/0004-6361/201220822}, \href {https://ui.adsabs.harvard.edu/abs/2013A&A...553A.119A} {553, A119}

\bibitem[\protect\citeauthoryear{{Arzoumanian} et~al.,}{{Arzoumanian} et~al.}{2019}]{2019A&A...621A..42A}
{Arzoumanian} D.,  et~al., 2019, \mn@doi [\aap] {10.1051/0004-6361/201832725}, \href {https://ui.adsabs.harvard.edu/abs/2019A&A...621A..42A} {621, A42}

\bibitem[\protect\citeauthoryear{{Bally}}{{Bally}}{2016}]{2016ARA&A..54..491B}
{Bally} J.,  2016, \mn@doi [\araa] {10.1146/annurev-astro-081915-023341}, \href {https://ui.adsabs.harvard.edu/abs/2016ARA&A..54..491B} {54, 491}

\bibitem[\protect\citeauthoryear{Barnes, Muller, Indermuehle, O'Dougherty, Lowe, Cunningham, Hernandez  \& Fuller}{Barnes et~al.}{2015}]{Barnes_2015}
Barnes P.~J.,  Muller E.,  Indermuehle B.,  O'Dougherty S.~N.,  Lowe V.,  Cunningham M.,  Hernandez A.~K.,   Fuller G.~A.,  2015, \mn@doi [The Astrophysical Journal] {10.1088/0004-637x/812/1/6}, 812, 6

\bibitem[\protect\citeauthoryear{{Benjamin} et~al.,}{{Benjamin} et~al.}{2003}]{2003PASP..115..953B}
{Benjamin} R.~A.,  et~al., 2003, \mn@doi [\pasp] {10.1086/376696}, \href {https://ui.adsabs.harvard.edu/abs/2003PASP..115..953B} {115, 953}

\bibitem[\protect\citeauthoryear{{Beuther}, {Walsh}, {Schilke}, {Sridharan}, {Menten}  \& {Wyrowski}}{{Beuther} et~al.}{2002}]{2002A&A...390..289B}
{Beuther} H.,  {Walsh} A.,  {Schilke} P.,  {Sridharan} T.~K.,  {Menten} K.~M.,   {Wyrowski} F.,  2002, \mn@doi [\aap] {10.1051/0004-6361:20020710}, \href {https://ui.adsabs.harvard.edu/abs/2002A&A...390..289B} {390, 289}

\bibitem[\protect\citeauthoryear{{Caswell}}{{Caswell}}{2003}]{2003MNRAS.341..551C}
{Caswell} J.~L.,  2003, \mn@doi [\mnras] {10.1046/j.1365-8711.2003.06418.x}, \href {https://ui.adsabs.harvard.edu/abs/2003MNRAS.341..551C} {341, 551}

\bibitem[\protect\citeauthoryear{{Caswell} \& {Vaile}}{{Caswell} \& {Vaile}}{1995}]{1995MNRAS.273..328C}
{Caswell} J.~L.,  {Vaile} R.~A.,  1995, \mn@doi [\mnras] {10.1093/mnras/273.2.328}, \href {https://ui.adsabs.harvard.edu/abs/1995MNRAS.273..328C} {273, 328}

\bibitem[\protect\citeauthoryear{{Caswell}, {Batchelor}, {Forster}  \& {Wellington}}{{Caswell} et~al.}{1983}]{1983AuJPh..36..401C}
{Caswell} J.~L.,  {Batchelor} R.~A.,  {Forster} J.~R.,   {Wellington} K.~J.,  1983, \mn@doi [Australian Journal of Physics] {10.1071/PH830401b}, \href {https://ui.adsabs.harvard.edu/abs/1983AuJPh..36..401C} {36, 401}

\bibitem[\protect\citeauthoryear{{Caswell}, {Vaile}, {Ellingsen}, {Whiteoak}  \& {Norris}}{{Caswell} et~al.}{1995}]{1995MNRAS.272...96C}
{Caswell} J.~L.,  {Vaile} R.~A.,  {Ellingsen} S.~P.,  {Whiteoak} J.~B.,   {Norris} R.~P.,  1995, \mn@doi [\mnras] {10.1093/mnras/272.1.96}, \href {https://ui.adsabs.harvard.edu/abs/1995MNRAS.272...96C} {272, 96}

\bibitem[\protect\citeauthoryear{{Chambers}, {Jackson}, {Rathborne}  \& {Simon}}{{Chambers} et~al.}{2009}]{2009ApJS..181..360C}
{Chambers} E.~T.,  {Jackson} J.~M.,  {Rathborne} J.~M.,   {Simon} R.,  2009, \mn@doi [\apjs] {10.1088/0067-0049/181/2/360}, \href {https://ui.adsabs.harvard.edu/abs/2009ApJS..181..360C} {181, 360}

\bibitem[\protect\citeauthoryear{{Churchwell}}{{Churchwell}}{2002}]{2002ARA&A..40...27C}
{Churchwell} E.,  2002, \mn@doi [\araa] {10.1146/annurev.astro.40.060401.093845}, \href {https://ui.adsabs.harvard.edu/abs/2002ARA&A..40...27C} {40, 27}

\bibitem[\protect\citeauthoryear{{Clarke}, {Whitworth}  \& {Hubber}}{{Clarke} et~al.}{2016}]{2016MNRAS.458..319C}
{Clarke} S.~D.,  {Whitworth} A.~P.,   {Hubber} D.~A.,  2016, \mn@doi [\mnras] {10.1093/mnras/stw407}, \href {https://ui.adsabs.harvard.edu/abs/2016MNRAS.458..319C} {458, 319}

\bibitem[\protect\citeauthoryear{{Cohen}, {Masheder}  \& {Caswell}}{{Cohen} et~al.}{1995}]{1995MNRAS.274..808C}
{Cohen} R.~J.,  {Masheder} M.~R.~W.,   {Caswell} J.~L.,  1995, \mn@doi [\mnras] {10.1093/mnras/274.3.808}, \href {https://ui.adsabs.harvard.edu/abs/1995MNRAS.274..808C} {274, 808}

\bibitem[\protect\citeauthoryear{{Didelon} et~al.,}{{Didelon} et~al.}{2015}]{2015A&A...584A...4D}
{Didelon} P.,  et~al., 2015, \mn@doi [\aap] {10.1051/0004-6361/201526239}, \href {https://ui.adsabs.harvard.edu/abs/2015A&A...584A...4D} {584, A4}

\bibitem[\protect\citeauthoryear{{Draine}}{{Draine}}{2003}]{2003ARA&A..41..241D}
{Draine} B.~T.,  2003, \mn@doi [\araa] {10.1146/annurev.astro.41.011802.094840}, \href {https://ui.adsabs.harvard.edu/abs/2003ARA&A..41..241D} {41, 241}

\bibitem[\protect\citeauthoryear{{Duarte-Cabral}, {Chrysostomou}, {Peretto}, {Fuller}, {Matthews}, {Schieven}  \& {Davis}}{{Duarte-Cabral} et~al.}{2012}]{2012A&A...543A.140D}
{Duarte-Cabral} A.,  {Chrysostomou} A.,  {Peretto} N.,  {Fuller} G.~A.,  {Matthews} B.,  {Schieven} G.,   {Davis} G.~R.,  2012, \mn@doi [\aap] {10.1051/0004-6361/201219240}, \href {https://ui.adsabs.harvard.edu/abs/2012A&A...543A.140D} {543, A140}

\bibitem[\protect\citeauthoryear{{Flaherty}, {Pipher}, {Megeath}, {Winston}, {Gutermuth}, {Muzerolle}, {Allen}  \& {Fazio}}{{Flaherty} et~al.}{2007}]{2007ApJ...663.1069F}
{Flaherty} K.~M.,  {Pipher} J.~L.,  {Megeath} S.~T.,  {Winston} E.~M.,  {Gutermuth} R.~A.,  {Muzerolle} J.,  {Allen} L.~E.,   {Fazio} G.~G.,  2007, \mn@doi [\apj] {10.1086/518411}, \href {https://ui.adsabs.harvard.edu/abs/2007ApJ...663.1069F} {663, 1069}

\bibitem[\protect\citeauthoryear{{Foster} et~al.,}{{Foster} et~al.}{2013}]{2013PASA...30...38F}
{Foster} J.~B.,  et~al., 2013, \mn@doi [\pasa] {10.1017/pasa.2013.18}, \href {https://ui.adsabs.harvard.edu/abs/2013PASA...30...38F} {30, e038}

\bibitem[\protect\citeauthoryear{{Frank} et~al.,}{{Frank} et~al.}{2014}]{2014prpl.conf..451F}
{Frank} A.,  et~al., 2014, in {Beuther} H.,  {Klessen} R.~S.,  {Dullemond} C.~P.,   {Henning} T.,  eds, Protostars and Planets VI. pp 451--474 (\mn@eprint {arXiv} {1402.3553}), \mn@doi{10.2458/azu_uapress_9780816531240-ch020}

\bibitem[\protect\citeauthoryear{{Fukui} et~al.,}{{Fukui} et~al.}{2018}]{2018ApJ...859..166F}
{Fukui} Y.,  et~al., 2018, \mn@doi [\apj] {10.3847/1538-4357/aac217}, \href {https://ui.adsabs.harvard.edu/abs/2018ApJ...859..166F} {859, 166}

\bibitem[\protect\citeauthoryear{{Garc{\'\i}a}, {Bronfman}, {Nyman}, {Dame}  \& {Luna}}{{Garc{\'\i}a} et~al.}{2014}]{2014ApJS..212....2G}
{Garc{\'\i}a} P.,  {Bronfman} L.,  {Nyman} L.-{\r{A}}.,  {Dame} T.~M.,   {Luna} A.,  2014, \mn@doi [\apjs] {10.1088/0067-0049/212/1/2}, \href {https://ui.adsabs.harvard.edu/abs/2014ApJS..212....2G} {212, 2}

\bibitem[\protect\citeauthoryear{{Ginard} et~al.,}{{Ginard} et~al.}{2012}]{2012A&A...543A..27G}
{Ginard} D.,  et~al., 2012, \mn@doi [\aap] {10.1051/0004-6361/201118347}, \href {https://ui.adsabs.harvard.edu/abs/2012A&A...543A..27G} {543, A27}

\bibitem[\protect\citeauthoryear{{Gregersen}, {Evans}, {Zhou}  \& {Choi}}{{Gregersen} et~al.}{1997}]{1997ApJ...484..256G}
{Gregersen} E.~M.,  {Evans} Neal~J. I.,  {Zhou} S.,   {Choi} M.,  1997, \mn@doi [\apj] {10.1086/304297}, \href {https://ui.adsabs.harvard.edu/abs/1997ApJ...484..256G} {484, 256}

\bibitem[\protect\citeauthoryear{{Guzm{\'a}n}, {May}, {Alvarez}  \& {Maeda}}{{Guzm{\'a}n} et~al.}{2011}]{2011A&A...525A.138G}
{Guzm{\'a}n} A.~E.,  {May} J.,  {Alvarez} H.,   {Maeda} K.,  2011, \mn@doi [\aap] {10.1051/0004-6361/200913628}, \href {https://ui.adsabs.harvard.edu/abs/2011A&A...525A.138G} {525, A138}

\bibitem[\protect\citeauthoryear{{Hacar}, {Tafalla}, {Forbrich}, {Alves}, {Meingast}, {Grossschedl}  \& {Teixeira}}{{Hacar} et~al.}{2018}]{2018A&A...610A..77H}
{Hacar} A.,  {Tafalla} M.,  {Forbrich} J.,  {Alves} J.,  {Meingast} S.,  {Grossschedl} J.,   {Teixeira} P.~S.,  2018, \mn@doi [\aap] {10.1051/0004-6361/201731894}, \href {https://ui.adsabs.harvard.edu/abs/2018A&A...610A..77H} {610, A77}

\bibitem[\protect\citeauthoryear{{Hartmann}, {Megeath}, {Allen}, {Luhman}, {Calvet}, {D'Alessio}, {Franco-Hernandez}  \& {Fazio}}{{Hartmann} et~al.}{2005}]{2005ApJ...629..881H}
{Hartmann} L.,  {Megeath} S.~T.,  {Allen} L.,  {Luhman} K.,  {Calvet} N.,  {D'Alessio} P.,  {Franco-Hernandez} R.,   {Fazio} G.,  2005, \mn@doi [\apj] {10.1086/431472}, \href {https://ui.adsabs.harvard.edu/abs/2005ApJ...629..881H} {629, 881}

\bibitem[\protect\citeauthoryear{{Haslam}, {Salter}, {Stoffel}  \& {Wilson}}{{Haslam} et~al.}{1982}]{1982A&AS...47....1H}
{Haslam} C.~G.~T.,  {Salter} C.~J.,  {Stoffel} H.,   {Wilson} W.~E.,  1982, \aaps, \href {https://ui.adsabs.harvard.edu/abs/1982A&AS...47....1H} {47, 1}

\bibitem[\protect\citeauthoryear{{Heigl}, {Hoemann}  \& {Burkert}}{{Heigl} et~al.}{2022}]{2022MNRAS.517.5272H}
{Heigl} S.,  {Hoemann} E.,   {Burkert} A.,  2022, \mn@doi [\mnras] {10.1093/mnras/stac3110}, \href {https://ui.adsabs.harvard.edu/abs/2022MNRAS.517.5272H} {517, 5272}

\bibitem[\protect\citeauthoryear{{Inutsuka}, {Inoue}, {Iwasaki}  \& {Hosokawa}}{{Inutsuka} et~al.}{2015}]{2015A&A...580A..49I}
{Inutsuka} S.-i.,  {Inoue} T.,  {Iwasaki} K.,   {Hosokawa} T.,  2015, \mn@doi [\aap] {10.1051/0004-6361/201425584}, \href {https://ui.adsabs.harvard.edu/abs/2015A&A...580A..49I} {580, A49}

\bibitem[\protect\citeauthoryear{{Jackson}, {Finn}, {Chambers}, {Rathborne}  \& {Simon}}{{Jackson} et~al.}{2010}]{2010ApJ...719L.185J}
{Jackson} J.~M.,  {Finn} S.~C.,  {Chambers} E.~T.,  {Rathborne} J.~M.,   {Simon} R.,  2010, \mn@doi [\apjl] {10.1088/2041-8205/719/2/L185}, \href {https://ui.adsabs.harvard.edu/abs/2010ApJ...719L.185J} {719, L185}

\bibitem[\protect\citeauthoryear{{Jackson} et~al.,}{{Jackson} et~al.}{2013}]{2013PASA...30...57J}
{Jackson} J.~M.,  et~al., 2013, \mn@doi [\pasa] {10.1017/pasa.2013.37}, \href {https://ui.adsabs.harvard.edu/abs/2013PASA...30...57J} {30, e057}

\bibitem[\protect\citeauthoryear{{Juvela} et~al.,}{{Juvela} et~al.}{2012}]{2012A&A...541A..12J}
{Juvela} M.,  et~al., 2012, \mn@doi [\aap] {10.1051/0004-6361/201118640}, \href {https://ui.adsabs.harvard.edu/abs/2012A&A...541A..12J} {541, A12}

\bibitem[\protect\citeauthoryear{{Kirk}, {Myers}, {Bourke}, {Gutermuth}, {Hedden}  \& {Wilson}}{{Kirk} et~al.}{2013}]{2013ApJ...766..115K}
{Kirk} H.,  {Myers} P.~C.,  {Bourke} T.~L.,  {Gutermuth} R.~A.,  {Hedden} A.,   {Wilson} G.~W.,  2013, \mn@doi [\apj] {10.1088/0004-637X/766/2/115}, \href {https://ui.adsabs.harvard.edu/abs/2013ApJ...766..115K} {766, 115}

\bibitem[\protect\citeauthoryear{{K{\"o}nyves} et~al.,}{{K{\"o}nyves} et~al.}{2010}]{2010A&A...518L.106K}
{K{\"o}nyves} V.,  et~al., 2010, \mn@doi [\aap] {10.1051/0004-6361/201014689}, \href {https://ui.adsabs.harvard.edu/abs/2010A&A...518L.106K} {518, L106}

\bibitem[\protect\citeauthoryear{{K{\"o}nyves} et~al.,}{{K{\"o}nyves} et~al.}{2015}]{2015A&A...584A..91K}
{K{\"o}nyves} V.,  et~al., 2015, \mn@doi [\aap] {10.1051/0004-6361/201525861}, \href {https://ui.adsabs.harvard.edu/abs/2015A&A...584A..91K} {584, A91}

\bibitem[\protect\citeauthoryear{{Kuffmeier}, {Calcutt}  \& {Kristensen}}{{Kuffmeier} et~al.}{2019}]{2019A&A...628A.112K}
{Kuffmeier} M.,  {Calcutt} H.,   {Kristensen} L.~E.,  2019, \mn@doi [\aap] {10.1051/0004-6361/201935504}, \href {https://ui.adsabs.harvard.edu/abs/2019A&A...628A.112K} {628, A112}

\bibitem[\protect\citeauthoryear{{Kumar}, {Arzoumanian}, {Men'shchikov}, {Palmeirim}, {Matsumura}  \& {Inutsuka}}{{Kumar} et~al.}{2022}]{2022A&A...658A.114K}
{Kumar} M.~S.~N.,  {Arzoumanian} D.,  {Men'shchikov} A.,  {Palmeirim} P.,  {Matsumura} M.,   {Inutsuka} S.,  2022, \mn@doi [\aap] {10.1051/0004-6361/202140363}, \href {https://ui.adsabs.harvard.edu/abs/2022A&A...658A.114K} {658, A114}

\bibitem[\protect\citeauthoryear{{Lada} \& {Lada}}{{Lada} \& {Lada}}{2003}]{2003ARA&A..41...57L}
{Lada} C.~J.,  {Lada} E.~A.,  2003, \mn@doi [\araa] {10.1146/annurev.astro.41.011802.094844}, \href {https://ui.adsabs.harvard.edu/abs/2003ARA&A..41...57L} {41, 57}

\bibitem[\protect\citeauthoryear{{Liu}, {Jim{\'e}nez-Serra}, {Ho}, {Chen}, {Zhang}  \& {Li}}{{Liu} et~al.}{2012}]{2012ApJ...756...10L}
{Liu} H.~B.,  {Jim{\'e}nez-Serra} I.,  {Ho} P. T.~P.,  {Chen} H.-R.,  {Zhang} Q.,   {Li} Z.-Y.,  2012, \mn@doi [\apj] {10.1088/0004-637X/756/1/10}, \href {https://ui.adsabs.harvard.edu/abs/2012ApJ...756...10L} {756, 10}

\bibitem[\protect\citeauthoryear{{Liu}, {Xu}, {Wang}, {Yu}, {Zhang}, {Li}  \& {Zhang}}{{Liu} et~al.}{2021}]{2021A&A...646A.137L}
{Liu} X.-L.,  {Xu} J.-L.,  {Wang} J.-J.,  {Yu} N.-P.,  {Zhang} C.-P.,  {Li} N.,   {Zhang} G.-Y.,  2021, \mn@doi [\aap] {10.1051/0004-6361/201935035}, \href {https://ui.adsabs.harvard.edu/abs/2021A&A...646A.137L} {646, A137}

\bibitem[\protect\citeauthoryear{{Low} et~al.,}{{Low} et~al.}{1984}]{1984ApJ...278L..19L}
{Low} F.~J.,  et~al., 1984, \mn@doi [\apjl] {10.1086/184213}, \href {https://ui.adsabs.harvard.edu/abs/1984ApJ...278L..19L} {278, L19}

\bibitem[\protect\citeauthoryear{{Mac Low}}{{Mac Low}}{2008}]{2008ASPC..387..148M}
{Mac Low} M.~M.,  2008, in {Beuther} H.,  {Linz} H.,   {Henning} T.,  eds,  Astronomical Society of the Pacific Conference Series Vol. 387, Massive Star Formation: Observations Confront Theory. pp 148--157

\bibitem[\protect\citeauthoryear{{Maity}, {Dewangan}, {Sano}, {Tachihara}, {Fukui}  \& {Bhadari}}{{Maity} et~al.}{2022}]{2022ApJ...934....2M}
{Maity} A.~K.,  {Dewangan} L.~K.,  {Sano} H.,  {Tachihara} K.,  {Fukui} Y.,   {Bhadari} N.~K.,  2022, \mn@doi [\apj] {10.3847/1538-4357/ac7872}, \href {https://ui.adsabs.harvard.edu/abs/2022ApJ...934....2M} {934, 2}

\bibitem[\protect\citeauthoryear{{Mardones}, {Myers}, {Tafalla}, {Wilner}, {Bachiller}  \& {Garay}}{{Mardones} et~al.}{1997}]{1997ApJ...489..719M}
{Mardones} D.,  {Myers} P.~C.,  {Tafalla} M.,  {Wilner} D.~J.,  {Bachiller} R.,   {Garay} G.,  1997, \mn@doi [\apj] {10.1086/304812}, \href {https://ui.adsabs.harvard.edu/abs/1997ApJ...489..719M} {489, 719}

\bibitem[\protect\citeauthoryear{{Martins}, {Schaerer}  \& {Hillier}}{{Martins} et~al.}{2005}]{2005A&A...436.1049M}
{Martins} F.,  {Schaerer} D.,   {Hillier} D.~J.,  2005, \mn@doi [\aap] {10.1051/0004-6361:20042386}, \href {https://ui.adsabs.harvard.edu/abs/2005A&A...436.1049M} {436, 1049}

\bibitem[\protect\citeauthoryear{{Megeath} et~al.,}{{Megeath} et~al.}{2004}]{2004ApJS..154..367M}
{Megeath} S.~T.,  et~al., 2004, \mn@doi [\apjs] {10.1086/422823}, \href {https://ui.adsabs.harvard.edu/abs/2004ApJS..154..367M} {154, 367}

\bibitem[\protect\citeauthoryear{{Men'shchikov}}{{Men'shchikov}}{2016}]{2016A&A...593A..71M}
{Men'shchikov} A.,  2016, \mn@doi [\aap] {10.1051/0004-6361/201628122}, \href {https://ui.adsabs.harvard.edu/abs/2016A&A...593A..71M} {593, A71}

\bibitem[\protect\citeauthoryear{{Men'shchikov}}{{Men'shchikov}}{2021a}]{2021A&A...649A..89M}
{Men'shchikov} A.,  2021a, \mn@doi [\aap] {10.1051/0004-6361/202039913}, \href {https://ui.adsabs.harvard.edu/abs/2021A&A...649A..89M} {649, A89}

\bibitem[\protect\citeauthoryear{{Men'shchikov}}{{Men'shchikov}}{2021b}]{2021A&A...654A..78M}
{Men'shchikov} A.,  2021b, \mn@doi [\aap] {10.1051/0004-6361/202141533}, \href {https://ui.adsabs.harvard.edu/abs/2021A&A...654A..78M} {654, A78}

\bibitem[\protect\citeauthoryear{{Men'shchikov} et~al.,}{{Men'shchikov} et~al.}{2010}]{2010A&A...518L.103M}
{Men'shchikov} A.,  et~al., 2010, \mn@doi [\aap] {10.1051/0004-6361/201014668}, \href {https://ui.adsabs.harvard.edu/abs/2010A&A...518L.103M} {518, L103}

\bibitem[\protect\citeauthoryear{{Menten}}{{Menten}}{1991}]{1991ASPC...16..119M}
{Menten} K.~M.,  1991, in {Haschick} A.~D.,  {Ho} P. T.~P.,  eds,  Astronomical Society of the Pacific Conference Series Vol. 16, Atoms, Ions and Molecules: New Results in Spectral Line Astrophysics. pp 119--136

\bibitem[\protect\citeauthoryear{{Molinari} et~al.,}{{Molinari} et~al.}{2010a}]{2010PASP..122..314M}
{Molinari} S.,  et~al., 2010a, \mn@doi [\pasp] {10.1086/651314}, \href {https://ui.adsabs.harvard.edu/abs/2010PASP..122..314M} {122, 314}

\bibitem[\protect\citeauthoryear{{Molinari} et~al.,}{{Molinari} et~al.}{2010b}]{2010A&A...518L.100M}
{Molinari} S.,  et~al., 2010b, \mn@doi [\aap] {10.1051/0004-6361/201014659}, \href {https://ui.adsabs.harvard.edu/abs/2010A&A...518L.100M} {518, L100}

\bibitem[\protect\citeauthoryear{{Motte}, {Bontemps}  \& {Louvet}}{{Motte} et~al.}{2018}]{2018ARA&A..56...41M}
{Motte} F.,  {Bontemps} S.,   {Louvet} F.,  2018, \mn@doi [\araa] {10.1146/annurev-astro-091916-055235}, \href {https://ui.adsabs.harvard.edu/abs/2018ARA&A..56...41M} {56, 41}

\bibitem[\protect\citeauthoryear{{Myers}}{{Myers}}{2009}]{2009ApJ...700.1609M}
{Myers} P.~C.,  2009, \mn@doi [\apj] {10.1088/0004-637X/700/2/1609}, \href {https://ui.adsabs.harvard.edu/abs/2009ApJ...700.1609M} {700, 1609}

\bibitem[\protect\citeauthoryear{{Noriega-Crespo} et~al.,}{{Noriega-Crespo} et~al.}{2004}]{2004ApJS..154..352N}
{Noriega-Crespo} A.,  et~al., 2004, \mn@doi [\apjs] {10.1086/422819}, \href {https://ui.adsabs.harvard.edu/abs/2004ApJS..154..352N} {154, 352}

\bibitem[\protect\citeauthoryear{{Ostriker}}{{Ostriker}}{1964}]{1964ApJ...140.1529O}
{Ostriker} J.,  1964, \mn@doi [\apj] {10.1086/148057}, \href {https://ui.adsabs.harvard.edu/abs/1964ApJ...140.1529O} {140, 1529}

\bibitem[\protect\citeauthoryear{{Panopoulou}, {Clark}, {Hacar}, {Heitsch}, {Kainulainen}, {Ntormousi}, {Seifried}  \& {Smith}}{{Panopoulou} et~al.}{2022}]{2022A&A...657L..13P}
{Panopoulou} G.~V.,  {Clark} S.~E.,  {Hacar} A.,  {Heitsch} F.,  {Kainulainen} J.,  {Ntormousi} E.,  {Seifried} D.,   {Smith} R.~J.,  2022, \mn@doi [\aap] {10.1051/0004-6361/202142281}, \href {https://ui.adsabs.harvard.edu/abs/2022A&A...657L..13P} {657, L13}

\bibitem[\protect\citeauthoryear{{Peretto} et~al.,}{{Peretto} et~al.}{2014}]{2014A&A...561A..83P}
{Peretto} N.,  et~al., 2014, \mn@doi [\aap] {10.1051/0004-6361/201322172}, \href {https://ui.adsabs.harvard.edu/abs/2014A&A...561A..83P} {561, A83}

\bibitem[\protect\citeauthoryear{{Peretto}, {Rigby}, {Louvet}, {Fuller}, {Traficante}  \& {Gaudel}}{{Peretto} et~al.}{2023}]{2023MNRAS.525.2935P}
{Peretto} N.,  {Rigby} A.~J.,  {Louvet} F.,  {Fuller} G.~A.,  {Traficante} A.,   {Gaudel} M.,  2023, \mn@doi [\mnras] {10.1093/mnras/stad2453}, \href {https://ui.adsabs.harvard.edu/abs/2023MNRAS.525.2935P} {525, 2935}

\bibitem[\protect\citeauthoryear{{Pirogov}, {Zinchenko}, {Caselli}, {Johansson}  \& {Myers}}{{Pirogov} et~al.}{2003}]{2003A&A...405..639P}
{Pirogov} L.,  {Zinchenko} I.,  {Caselli} P.,  {Johansson} L.~E.~B.,   {Myers} P.~C.,  2003, \mn@doi [\aap] {10.1051/0004-6361:20030659}, \href {https://ui.adsabs.harvard.edu/abs/2003A&A...405..639P} {405, 639}

\bibitem[\protect\citeauthoryear{{Polderman}, {Haverkorn}  \& {Jaffe}}{{Polderman} et~al.}{2020}]{2020A&A...636A...2P}
{Polderman} I.~M.,  {Haverkorn} M.,   {Jaffe} T.~R.,  2020, \mn@doi [\aap] {10.1051/0004-6361/201937042}, \href {https://ui.adsabs.harvard.edu/abs/2020A&A...636A...2P} {636, A2}

\bibitem[\protect\citeauthoryear{{Polychroni} et~al.,}{{Polychroni} et~al.}{2013}]{2013ApJ...777L..33P}
{Polychroni} D.,  et~al., 2013, \mn@doi [\apjl] {10.1088/2041-8205/777/2/L33}, \href {https://ui.adsabs.harvard.edu/abs/2013ApJ...777L..33P} {777, L33}

\bibitem[\protect\citeauthoryear{{Ragan}, {Heitsch}, {Bergin}  \& {Wilner}}{{Ragan} et~al.}{2012}]{2012ApJ...746..174R}
{Ragan} S.~E.,  {Heitsch} F.,  {Bergin} E.~A.,   {Wilner} D.,  2012, \mn@doi [\apj] {10.1088/0004-637X/746/2/174}, \href {https://ui.adsabs.harvard.edu/abs/2012ApJ...746..174R} {746, 174}

\bibitem[\protect\citeauthoryear{{Ragan}, {Henning}, {Tackenberg}, {Beuther}, {Johnston}, {Kainulainen}  \& {Linz}}{{Ragan} et~al.}{2014}]{2014A&A...568A..73R}
{Ragan} S.~E.,  {Henning} T.,  {Tackenberg} J.,  {Beuther} H.,  {Johnston} K.~G.,  {Kainulainen} J.,   {Linz} H.,  2014, \mn@doi [\aap] {10.1051/0004-6361/201423401}, \href {https://ui.adsabs.harvard.edu/abs/2014A&A...568A..73R} {568, A73}

\bibitem[\protect\citeauthoryear{{Rathborne}, {Jackson}  \& {Simon}}{{Rathborne} et~al.}{2006}]{2006ApJ...641..389R}
{Rathborne} J.~M.,  {Jackson} J.~M.,   {Simon} R.,  2006, \mn@doi [\apj] {10.1086/500423}, \href {https://ui.adsabs.harvard.edu/abs/2006ApJ...641..389R} {641, 389}

\bibitem[\protect\citeauthoryear{Ray et~al.,}{Ray et~al.}{2023}]{Ray2023}
Ray T.~P.,  et~al., 2023, \mn@doi [Nature] {10.1038/s41586-023-06551-1}, 622, 48

\bibitem[\protect\citeauthoryear{{Reid} et~al.,}{{Reid} et~al.}{2014}]{2014ApJ...783..130R}
{Reid} M.~J.,  et~al., 2014, \mn@doi [\apj] {10.1088/0004-637X/783/2/130}, \href {https://ui.adsabs.harvard.edu/abs/2014ApJ...783..130R} {783, 130}

\bibitem[\protect\citeauthoryear{{Remazeilles}, {Dickinson}, {Banday}, {Bigot-Sazy}  \& {Ghosh}}{{Remazeilles} et~al.}{2015}]{2015MNRAS.451.4311R}
{Remazeilles} M.,  {Dickinson} C.,  {Banday} A.~J.,  {Bigot-Sazy} M.~A.,   {Ghosh} T.,  2015, \mn@doi [\mnras] {10.1093/mnras/stv1274}, \href {https://ui.adsabs.harvard.edu/abs/2015MNRAS.451.4311R} {451, 4311}

\bibitem[\protect\citeauthoryear{{Rodgers}, {Campbell}  \& {Whiteoak}}{{Rodgers} et~al.}{1960}]{1960MNRAS.121..103R}
{Rodgers} A.~W.,  {Campbell} C.~T.,   {Whiteoak} J.~B.,  1960, \mn@doi [\mnras] {10.1093/mnras/121.1.103}, \href {https://ui.adsabs.harvard.edu/abs/1960MNRAS.121..103R} {121, 103}

\bibitem[\protect\citeauthoryear{{Roger}, {Costain}, {Landecker}  \& {Swerdlyk}}{{Roger} et~al.}{1999}]{1999A&AS..137....7R}
{Roger} R.~S.,  {Costain} C.~H.,  {Landecker} T.~L.,   {Swerdlyk} C.~M.,  1999, \mn@doi [\aaps] {10.1051/aas:1999239}, \href {https://ui.adsabs.harvard.edu/abs/1999A&AS..137....7R} {137, 7}

\bibitem[\protect\citeauthoryear{{Rybicki} \& {Lightman}}{{Rybicki} \& {Lightman}}{1985}]{rybicki.lightman.ch5}
{Rybicki} G.~B.,  {Lightman} A.~P.,  1985, BREMSSTRAHLUNG.
John Wiley and Sons, Ltd, pp 155--166 (\mn@eprint {} {https://onlinelibrary.wiley.com/doi/pdf/10.1002/9783527618170.ch5}), \mn@doi{https://doi.org/10.1002/9783527618170.ch5}, \url {https://onlinelibrary.wiley.com/doi/abs/10.1002/9783527618170.ch5}

\bibitem[\protect\citeauthoryear{{Samal}, {Pandey}, {Ojha}, {Chauhan}, {Jose}  \& {Pandey}}{{Samal} et~al.}{2012}]{2012ApJ...755...20S}
{Samal} M.~R.,  {Pandey} A.~K.,  {Ojha} D.~K.,  {Chauhan} N.,  {Jose} J.,   {Pandey} B.,  2012, \mn@doi [\apj] {10.1088/0004-637X/755/1/20}, \href {https://ui.adsabs.harvard.edu/abs/2012ApJ...755...20S} {755, 20}

\bibitem[\protect\citeauthoryear{{Schmiedeke} et~al.,}{{Schmiedeke} et~al.}{2016}]{2016A&A...588A.143S}
{Schmiedeke} A.,  et~al., 2016, \mn@doi [\aap] {10.1051/0004-6361/201527311}, \href {https://ui.adsabs.harvard.edu/abs/2016A&A...588A.143S} {588, A143}

\bibitem[\protect\citeauthoryear{{Smith}, {Longmore}  \& {Bonnell}}{{Smith} et~al.}{2009}]{2009MNRAS.400.1775S}
{Smith} R.~J.,  {Longmore} S.,   {Bonnell} I.,  2009, \mn@doi [\mnras] {10.1111/j.1365-2966.2009.15621.x}, \href {https://ui.adsabs.harvard.edu/abs/2009MNRAS.400.1775S} {400, 1775}

\bibitem[\protect\citeauthoryear{{Sollins} \& {Megeath}}{{Sollins} \& {Megeath}}{2004}]{2004AJ....128.2374S}
{Sollins} P.~K.,  {Megeath} S.~T.,  2004, \mn@doi [\aj] {10.1086/425044}, \href {https://ui.adsabs.harvard.edu/abs/2004AJ....128.2374S} {128, 2374}

\bibitem[\protect\citeauthoryear{{Swarup}, {Ananthakrishnan}, {Kapahi}, {Rao}, {Subrahmanya}  \& {Kulkarni}}{{Swarup} et~al.}{1991}]{1991CSci...60...95S}
{Swarup} G.,  {Ananthakrishnan} S.,  {Kapahi} V.~K.,  {Rao} A.~P.,  {Subrahmanya} C.~R.,   {Kulkarni} V.~K.,  1991, Current Science, \href {https://ui.adsabs.harvard.edu/abs/1991CSci...60...95S} {60, 95}

\bibitem[\protect\citeauthoryear{{Tackenberg} et~al.,}{{Tackenberg} et~al.}{2014}]{2014A&A...565A.101T}
{Tackenberg} J.,  et~al., 2014, \mn@doi [\aap] {10.1051/0004-6361/201321555}, \href {https://ui.adsabs.harvard.edu/abs/2014A&A...565A.101T} {565, A101}

\bibitem[\protect\citeauthoryear{{Takami}, {Karr}, {Koh}, {Chen}  \& {Lee}}{{Takami} et~al.}{2010}]{2010ApJ...720..155T}
{Takami} M.,  {Karr} J.~L.,  {Koh} H.,  {Chen} H.-H.,   {Lee} H.-T.,  2010, \mn@doi [\apj] {10.1088/0004-637X/720/1/155}, \href {https://ui.adsabs.harvard.edu/abs/2010ApJ...720..155T} {720, 155}

\bibitem[\protect\citeauthoryear{{Tan}}{{Tan}}{2015}]{2015IAUGA..2254046T}
{Tan} J.~C.,  2015, in IAU General Assembly. p. 2254046

\bibitem[\protect\citeauthoryear{{Teixeira}, {McCoey}, {Fich}  \& {Lada}}{{Teixeira} et~al.}{2008}]{2008MNRAS.384...71T}
{Teixeira} P.~S.,  {McCoey} C.,  {Fich} M.,   {Lada} C.~J.,  2008, \mn@doi [\mnras] {10.1111/j.1365-2966.2007.12698.x}, \href {https://ui.adsabs.harvard.edu/abs/2008MNRAS.384...71T} {384, 71}

\bibitem[\protect\citeauthoryear{{Tobin}, {Hartmann}, {Bergin}, {Chiang}, {Looney}, {Chandler}, {Maret}  \& {Heitsch}}{{Tobin} et~al.}{2012}]{2012ApJ...748...16T}
{Tobin} J.~J.,  {Hartmann} L.,  {Bergin} E.,  {Chiang} H.-F.,  {Looney} L.~W.,  {Chandler} C.~J.,  {Maret} S.,   {Heitsch} F.,  2012, \mn@doi [\apj] {10.1088/0004-637X/748/1/16}, \href {https://ui.adsabs.harvard.edu/abs/2012ApJ...748...16T} {748, 16}

\bibitem[\protect\citeauthoryear{{Trevi{\~n}o-Morales} et~al.,}{{Trevi{\~n}o-Morales} et~al.}{2019}]{2019A&A...629A..81T}
{Trevi{\~n}o-Morales} S.~P.,  et~al., 2019, \mn@doi [\aap] {10.1051/0004-6361/201935260}, \href {https://ui.adsabs.harvard.edu/abs/2019A&A...629A..81T} {629, A81}

\bibitem[\protect\citeauthoryear{{V{\'a}zquez-Semadeni}, {Gonz{\'a}lez-Samaniego}  \& {Col{\'\i}n}}{{V{\'a}zquez-Semadeni} et~al.}{2017}]{2017MNRAS.467.1313V}
{V{\'a}zquez-Semadeni} E.,  {Gonz{\'a}lez-Samaniego} A.,   {Col{\'\i}n} P.,  2017, \mn@doi [\mnras] {10.1093/mnras/stw3229}, \href {https://ui.adsabs.harvard.edu/abs/2017MNRAS.467.1313V} {467, 1313}

\bibitem[\protect\citeauthoryear{{V{\'a}zquez-Semadeni}, {Palau}, {Ballesteros-Paredes}, {G{\'o}mez}  \& {Zamora-Avil{\'e}s}}{{V{\'a}zquez-Semadeni} et~al.}{2019}]{2019MNRAS.490.3061V}
{V{\'a}zquez-Semadeni} E.,  {Palau} A.,  {Ballesteros-Paredes} J.,  {G{\'o}mez} G.~C.,   {Zamora-Avil{\'e}s} M.,  2019, \mn@doi [\mnras] {10.1093/mnras/stz2736}, \href {https://ui.adsabs.harvard.edu/abs/2019MNRAS.490.3061V} {490, 3061}

\bibitem[\protect\citeauthoryear{{Velusamy}, {Peng}, {Li}, {Goldsmith}  \& {Langer}}{{Velusamy} et~al.}{2008}]{2008ApJ...688L..87V}
{Velusamy} T.,  {Peng} R.,  {Li} D.,  {Goldsmith} P.~F.,   {Langer} W.~D.,  2008, \mn@doi [\apjl] {10.1086/595585}, \href {https://ui.adsabs.harvard.edu/abs/2008ApJ...688L..87V} {688, L87}

\bibitem[\protect\citeauthoryear{{Wang}, {Koch}, {Galv{\'a}n-Madrid}, {Lai}, {Liu}, {Lin}  \& {Pattle}}{{Wang} et~al.}{2020}]{2020ApJ...905..158W}
{Wang} J.-W.,  {Koch} P.~M.,  {Galv{\'a}n-Madrid} R.,  {Lai} S.-P.,  {Liu} H.~B.,  {Lin} S.-J.,   {Pattle} K.,  2020, \mn@doi [\apj] {10.3847/1538-4357/abc74e}, \href {https://ui.adsabs.harvard.edu/abs/2020ApJ...905..158W} {905, 158}

\bibitem[\protect\citeauthoryear{{Wang} et~al.,}{{Wang} et~al.}{2022}]{2022ApJ...931..115W}
{Wang} J.-W.,  et~al., 2022, \mn@doi [\apj] {10.3847/1538-4357/ac6872}, \href {https://ui.adsabs.harvard.edu/abs/2022ApJ...931..115W} {931, 115}

\bibitem[\protect\citeauthoryear{{Wareing}, {Falle}  \& {Pittard}}{{Wareing} et~al.}{2019}]{2019MNRAS.485.4686W}
{Wareing} C.~J.,  {Falle} S.~A.~E.~G.,   {Pittard} J.~M.,  2019, \mn@doi [\mnras] {10.1093/mnras/stz768}, \href {https://ui.adsabs.harvard.edu/abs/2019MNRAS.485.4686W} {485, 4686}

\bibitem[\protect\citeauthoryear{{Werner} et~al.,}{{Werner} et~al.}{2004}]{2004ApJS..154....1W}
{Werner} M.~W.,  et~al., 2004, \mn@doi [\apjs] {10.1086/422992}, \href {https://ui.adsabs.harvard.edu/abs/2004ApJS..154....1W} {154, 1}

\bibitem[\protect\citeauthoryear{{Whitaker}, {Jackson}, {Rathborne}, {Foster}, {Contreras}, {Sanhueza}, {Stephens}  \& {Longmore}}{{Whitaker} et~al.}{2017}]{2017AJ....154..140W}
{Whitaker} J.~S.,  {Jackson} J.~M.,  {Rathborne} J.~M.,  {Foster} J.~B.,  {Contreras} Y.,  {Sanhueza} P.,  {Stephens} I.~W.,   {Longmore} S.~N.,  2017, \mn@doi [\aj] {10.3847/1538-3881/aa86ad}, \href {https://ui.adsabs.harvard.edu/abs/2017AJ....154..140W} {154, 140}

\bibitem[\protect\citeauthoryear{{Whitworth}}{{Whitworth}}{1979}]{1979MNRAS.186...59W}
{Whitworth} A.,  1979, \mn@doi [\mnras] {10.1093/mnras/186.1.59}, \href {https://ui.adsabs.harvard.edu/abs/1979MNRAS.186...59W} {186, 59}

\bibitem[\protect\citeauthoryear{{Wienen} et~al.,}{{Wienen} et~al.}{2015}]{2015A&A...579A..91W}
{Wienen} M.,  et~al., 2015, \mn@doi [\aap] {10.1051/0004-6361/201424802}, \href {https://ui.adsabs.harvard.edu/abs/2015A&A...579A..91W} {579, A91}

\bibitem[\protect\citeauthoryear{{Yuan} et~al.,}{{Yuan} et~al.}{2020}]{2020A&A...637A..67Y}
{Yuan} L.,  et~al., 2020, \mn@doi [\aap] {10.1051/0004-6361/201936625}, \href {https://ui.adsabs.harvard.edu/abs/2020A&A...637A..67Y} {637, A67}

\bibitem[\protect\citeauthoryear{{Zernickel}, {Schilke}  \& {Smith}}{{Zernickel} et~al.}{2013}]{2013A&A...554L...2Z}
{Zernickel} A.,  {Schilke} P.,   {Smith} R.~J.,  2013, \mn@doi [\aap] {10.1051/0004-6361/201321425}, \href {https://ui.adsabs.harvard.edu/abs/2013A&A...554L...2Z} {554, L2}

\bibitem[\protect\citeauthoryear{{Zhou} et~al.,}{{Zhou} et~al.}{2022}]{2022MNRAS.514.6038Z}
{Zhou} J.-W.,  et~al., 2022, \mn@doi [\mnras] {10.1093/mnras/stac1735}, \href {https://ui.adsabs.harvard.edu/abs/2022MNRAS.514.6038Z} {514, 6038}

\bibitem[\protect\citeauthoryear{{Zinnecker} \& {Yorke}}{{Zinnecker} \& {Yorke}}{2007}]{2007ARA&A..45..481Z}
{Zinnecker} H.,  {Yorke} H.~W.,  2007, \mn@doi [\araa] {10.1146/annurev.astro.44.051905.092549}, \href {https://ui.adsabs.harvard.edu/abs/2007ARA&A..45..481Z} {45, 481}

\makeatother
\end{thebibliography}



\appendix

\section{}


\begin{table*}
    \centering
	\caption{Properties of cores identified towards RCW~117. Here, D$_{eff}$ is the effective deconvolved size, N(H$_2$) is the peak column density and T$_{dust}$ is the associated dust temperature of the cores.}
	\label{tab:core_properties}
    \begin{tabular}{ccccccc} 
		\hline \hline
		No. & $\alpha_{\textrm{J}2000}$ & $\delta_{\textrm{J}2000}$ & D$_{eff}$~(pc) & N(H$_2$)~$\times~10^{22}$ (cm$^{-2}$) & T$_{dust}$~(K) & Mass~(M$_{\odot}$)\\
		\hline
		1 & 17:09:35.75 & -41:35:57.17 & 0.10 & 21.55 & 37.0 & 173.7 \\
		2 & 17:09:40.90 & -41:35:39.98 & 0.16 & 13.95 & 26.1 & 108.1 \\
		3 & 17:09:28.71 & -41:33:49.28 & 0.22 & 9.86 & 19.2 & 108.1 \\
		4 & 17:09:41.43 & -41:32:22.45 & 0.19 & 8.31 & 17.0 & 84.4 \\
		5 & 17:09:38.53 & -41:34:58.64 & 0.14 & 10.95 & 25.7 & 83.8 \\
		6 & 17:09:38.96 & -41:33:55.88 & 0.20 & 9.91 & 20.5 & 71.3 \\
		7 & 17:09:32.43 & -41:35:39.28 & 0.16 & 13.65 & 32.9 & 65.0 \\
		8 & 17:09:44.14 & -41:38:45.20 & 0.13 & 5.54 & 16.7 & 63.1 \\
		9 & 17:09:26.75 & -41:36:07.06 & 0.17 & 8.45 & 24.5 & 61.6 \\
		10 & 17:09:32.09 & -41:36:34.24 & 0.16 & 10.15 & 28.6 & 57.9 \\
		11 & 17:09:42.97 & -41:39:09.32 & 0.18 & 5.08 & 17.2 & 53.3 \\
		12 & 17:09:38.73 & -41:38:55.39 & 0.25 & 6.73 & 17.3 & 51.6 \\
		13 & 17:09:22.96 & -41:35:22.11 & 0.18 & 7.49 & 21.0 & 50.5 \\
		14 & 17:09:41.48 & -41:36:28.82 & 0.20 & 8.54 & 22.7 & 48.3 \\	
		15 & 17:09:24.97 & -41:34:06.72 & 0.26 & 4.37 & 19.7 & 47.1 \\
		16 & 17:09:33.66 & -41:36:03.15 & 0.26 & 10.71 & 33.4 & 41.5 \\
		17 & 17:09:55.10 & -41:35:31.98 & 0.16 & 3.93 & 18.9 & 33.1 \\
		18 & 17:09:38.15 & -41:33:14.69 & 0.14 & 4.67 & 19.8 & 31.2 \\
		19 & 17:09:43.69 & -41:38:19.65 & 0.18 & 4.45 & 17.2 & 29.8 \\
		20 & 17:09:08.37 & -41:40:02.40 & 0.17 & 5.23 & 16.7 & 29.3 \\
		21 & 17:09:32.50 & -41:28:58.35 & 0.18 & 3.82 & 16.9 & 27.0 \\
		22 & 17:09:39.71 & -41:32:57.72 & 0.16 & 3.48 & 19.3 & 26.9 \\
		23 & 17:09:31.37 & -41:33:51.56 & 0.16 & 5.25 & 24.7 & 26.1 \\
		24 & 17:09:10.13 & -41:39:49.98 & 0.21 & 4.29 & 17.6 & 25.6 \\
		25 & 17:09:43.39 & -41:38:00.49 & 0.16 & 4.18 & 18.1 & 25.1 \\
		26 & 17:09:02.96 & -41:43:11.02 & 0.17 & 3.62 & 16.5 & 21.6 \\
		27 & 17:09:45.60 & -41:39:20.23 & 0.15 & 3.15 & 17.6 & 21.1 \\
		28 & 17:09:29.08 & -41:37:12.42 & 0.17 & 5.16 & 23.6 & 21.0 \\
		29 & 17:09:35.15 & -41:30:07.83 & 0.16 & 2.11 & 17.5 & 19.1 \\
	    30 & 17:09:48.38 & -41:35:13.55 & 0.22 & 2.64 & 23.2 & 18.6 \\
		31 & 17:09:17.81 & -41:28:35.72 & 0.18 & 1.97 & 17.5 & 17.3 \\
		32 & 17:09:18.92 & -41:33:27.86 & 0.17 & 2.25 & 19.6 & 17.3 \\
		33 & 17:09:36.49 & -41:42:05.04 & 0.16 & 1.73 & 19.2 & 15.8 \\
		34 & 17:09:15.09 & -41:27:26.65 & 0.14 & 1.48 & 17.2 & 15.6 \\
		35 & 17:09:16.12 & -41:34:25.72 & 0.14 & 3.22 & 20.4 & 15.6 \\
		36 & 17:10:06.97 & -41:32:49.50 & 0.19 & 1.65 & 18.4 & 14.9 \\
		37 & 17:09:10.74 & -41:38:37.40 & 0.31 & 3.21 & 19.3 & 14.5 \\
		38 & 17:10:10.01 & -41:30:06.74 & 0.16 & 1.13 & 18.8 & 14.4 \\
		39 & 17:09:02.11 & -41:43:58.13 & 0.14 & 2.54 & 16.6 & 14.4 \\
		40 & 17:09:57.82 & -41:35:49.25 & 0.20 & 2.75 & 19.2 & 13.8 \\
        41 & 17:09:07.95 & -41:29:22.91 & 0.22 & 1.69 & 17.5 & 13.7 \\
		42 & 17:09:15.99 & -41:38:00.11 & 0.11 & 2.29 & 23.8 & 13.6 \\
		43 & 17:09:04.44 & -41:34:52.40 & 0.13 & 2.16 & 18.2 & 13.3 \\
		44 & 17:09:33.41 & -41:29:36.01 & 0.11 & 2.70 & 17.0 & 12.9 \\
		45 & 17:09:22.94 & -41:40:44.74 & 0.13 & 1.88 & 20.1 & 12.8 \\
		46 & 17:09:30.97 & -41:29:07.47 & 0.19 & 2.91 & 18.0 & 12.5 \\
		47 & 17:09:26.54 & -41:37:52.67 & 0.20 & 2.85 & 21.8 & 12.3 \\
		48 & 17:09:08.34 & -41:39:11.21 & 0.18 & 3.59 & 17.3 & 12.1 \\
		49 & 17:08:58.77 & -41:35:35.40 & 0.12 & 1.95 & 17.2 & 12.0 \\
		50 & 17:09:08.49 & -41:27:54.47 & 0.12 & 1.43 & 17.5 & 11.6 \\
		51 & 17:09:13.50 & -41:33:24.26 & 0.14 & 2.25 & 18.0 & 11.6 \\
		52 & 17:08:55.88 & -41:37:44.86 & 0.15 & 2.21 & 17.6 & 10.8 \\
		53 & 17:09:21.25 & -41:34:34.82 & 0.20 & 2.54 & 21.6 & 10.5 \\
		54 & 17:09:16.55 & -41:37:40.86 & 0.22 & 2.09 & 24.0 & 10.4 \\
		55 & 17:08:50.75 & -41:35:41.96 & 0.13 & 1.76 & 17.7 & 10.3 \\
		56 & 17:09:12.52 & -41:39:03.36 & 0.10 & 2.82 & 19.7 & 10.1 \\
		57 & 17:08:58.90 & -41:38:49.66 & 0.31 & 2.04 & 16.9 & 9.9 \\
		58 & 17:10:05.88 & -41:32:03.62 & 0.16 & 1.39 & 18.4 & 9.5 \\
		59 & 17:09:13.55 & -41:40:57.92 & 0.21 & 1.76 & 19.8 & 9.5 \\
		60 & 17:09:03.31 & -41:35:02.10 & 0.13 & 2.05 & 18.2 & 9.3 \\
		61 & 17:09:09.35 & -41:36:11.94 & 0.25 & 1.49 & 22.7 & 9.1 \\
		62 & 17:09:03.01 & -41:41:26.99 & 0.21 & 2.54 & 17.1 & 8.8 \\
		63 & 17:09:44.75 & -41:35:35.07 & 0.17 & 2.61 & 26.0 & 8.8 \\
		64 & 17:08:51.40 & -41:40:35.87 & 0.14 & 1.75 & 18.0 & 8.7 \\   	
		\hline
	\end{tabular}
\end{table*}


\begin{table*}
	\begin{tabular}{ccccccc} 
		\hline \hline
		No. & $\alpha_{\textrm{J}2000}$ & $\delta_{\textrm{J}2000}$ & D$_{eff}$~(pc) & N(H$_2$)~$\times~10^{22}$ (cm$^{-2}$) & T$_{dust}$~(K) & Mass~(M$_{\odot}$)\\
		\hline
		65 & 17:09:10.41 & -41:34:18.26 & 0.25 & 2.30 & 18.8 & 8.4 \\     
		66 & 17:09:27.36 & -41:28:00.64 & 0.16 & 1.91 & 17.5 & 8.2 \\
		67 & 17:09:31.81 & -41:31:50.57 & 0.21 & 1.08 & 23.7 & 7.8 \\
		68 & 17:09:29.14 & -41:41:21.98 & 0.12 & 1.44 & 20.4 & 7.4 \\
		69 & 17:09:43.81 & -41:31:29.02 & 0.23 & 1.73 & 17.8 & 7.3 \\
		70 & 17:09:10.95 & -41:36:54.99 & 0.15 & 1.78 & 25.8 & 7.0 \\
		71 & 17:09:26.27 & -41:29:09.55 & 0.14 & 1.78 & 18.3 & 6.9 \\
		72 & 17:09:12.04 & -41:32:32.52 & 0.12 & 1.66 & 18.3 & 6.9 \\
		73 & 17:09:27.42 & -41:27:19.57 & 0.19 & 1.89 & 17.3 & 6.7 \\
		74 & 17:09:45.65 & -41:42:57.79 & 0.14 & 1.29 & 18.4 & 6.5 \\
		75 & 17:09:12.50 & -41:28:57.02 & 0.15 & 1.39 & 17.9 & 6.4 \\
		76 & 17:09:39.64 & -41:30:54.94 & 0.14 & 1.30 & 18.4 & 6.3 \\
		77 & 17:09:00.95 & -41:34:25.19 & 0.12 & 1.67 & 17.6 & 6.3 \\
		78 & 17:09:25.38 & -41:40:29.57 & 0.21 & 1.44 & 20.5 & 6.2 \\
		79 & 17:09:30.62 & -41:42:24.22 & 0.12 & 1.00 & 20.3 & 5.6 \\
		80 & 17:09:02.70 & -41:37:26.13 & 0.17 & 1.66 & 18.9 & 5.6 \\
		81 & 17:08:51.89 & -41:34:40.68 & 0.21 & 1.42 & 18.2 & 5.5 \\
		82 & 17:09:18.98 & -41:30:22.27 & 0.14 & 0.80 & 19.4 & 5.2 \\
		83 & 17:09:29.37 & -41:43:04.34 & 0.17 & 0.83 & 20.1 & 5.1 \\
		84 & 17:10:00.14 & -41:36:26.57 & 0.27 & 1.52 & 19.6 & 4.5 \\
		85 & 17:09:29.68 & -41:32:00.39 & 0.17 & 0.98 & 23.1 & 3.9 \\
		86 & 17:10:13.10 & -41:29:32.49 & 0.16 & 0.73 & 18.7 & 2.9 \\
		87 & 17:09:47.84 & -41:27:10.45 & 0.17 & 0.59 & 20.1 & 2.5 \\
		88 & 17:10:02.52 & -41:33:07.11 & 0.07 & 0.72 & 20.5 & 1.7 \\
        \hline \hline
	\end{tabular}
\end{table*}


\begin{table*}
	\centering
	\caption{Details of Class 0/I YSOs identified towards RCW~117 from the \textit{Spitzer} colour-colour diagram.}
	\label{tab:classi_yso_table}
    \begin{tabular}{cccccccc} 
		\hline \hline
		No. & GLIMPSE designation & $\alpha_{\textrm{J}2000}$ & $\delta_{\textrm{J}2000}$ & $3.6~\mu$m & $4.5~\mu$m & $5.8~\mu$m & $8.0~\mu$m \\
         & & ($^{h}$ $^{m}$ $^{s}$) & ($^{\circ}$ $'$ $''$) & (mag) & (mag) & (mag) & (mag) \\
		\hline
        1 & SSTGLMA G345.3270-00.9886 & 17:09:29.67 & -41:41:06.71 & 13.88 $\pm$ 0.11 & 12.65 $\pm$ 0.11 & 11.28 $\pm$ 0.14 & - \\
        
        2 & SSTGLMA G345.3237-01.0057 & 17:09:33.46 & -41:41:52.97 & 14.75 $\pm$ 0.24 & 12.86 $\pm$ 0.13 & 11.77 $\pm$ 0.17 & 11.15 $\pm$ 0.18 \\
        
        3 & SSTGLMA G345.3206-00.9904 & 17:09:28.91 & -41:41:29.18 & 13.45 $\pm$ 0.09 & 12.19 $\pm$ 0.09 & 11.45 $\pm$ 0.15 & - \\
        
        4 & SSTGLMA G345.5321-01.0105 & 17:10:14.47 & -41:31:59.85 & 15.02 $\pm$ 0.21 & 13.91 $\pm$ 0.26 & 12.76 $\pm$ 0.30 & - \\
        
        5 & SSTGLMA G345.2951-00.9248 & 17:09:07.09 & -41:40:21.88 & 11.74 $\pm$ 0.09 & 10.36 $\pm$ 0.07 & 9.38 $\pm$ 0.06 & 9.44 $\pm$ 0.05 \\
        
        6 & SSTGLMA G345.2499-00.9427 & 17:09:03.06 & -41:43:11.27 & 8.65 $\pm$ 0.05 & 6.32 $\pm$ 0.07 & 4.64 $\pm$ 0.04 & 3.96 $\pm$ 0.07 \\
        
        7 & SSTGLMA G345.4495-01.0057 & 17:09:57.47 & -41:35:48.96 & 11.43 $\pm$ 0.05 & 10.04 $\pm$ 0.08 & 8.96 $\pm$ 0.04 & 7.89 $\pm$ 0.05 \\
        
        8 & SSTGLMA G345.4968-00.8788 & 17:09:33.78 & -41:29:00.03 & 14.87 $\pm$ 0.14 & 13.72 $\pm$ 0.16 & 12.39 $\pm$ 0.33 & 11.10 $\pm$ 0.13 \\
        
        9 & SSTGLMA G345.4683-00.9317 & 17:09:41.97 & -41:32:16.07 & 11.59 $\pm$ 0.04 & 10.23 $\pm$ 0.05 & 9.19 $\pm$ 0.05 & 9.38 $\pm$ 0.09 \\
        
        10 & SSTGLMA G345.4671-00.9322 & 17:09:41.88 & -41:32:20.50 & 12.55 $\pm$ 0.06 & 11.69 $\pm$ 0.08 & 10.91 $\pm$ 0.10 & 10.57 $\pm$ 0.08 \\
        
        11 & SSTGLMA G345.4639-00.9273 & 17:09:40.01 & -41:32:19.34 & 14.16 $\pm$ 0.17 & 13.41 $\pm$ 0.29 & 10.44 $\pm$ 0.12 & 9.14 $\pm$ 0.16 \\
        
        12 & SSTGLMA G345.4652-00.9312 & 17:09:41.26 & -41:32:23.91 & 14.15 $\pm$ 0.17 & 10.90 $\pm$ 0.07 & 9.42 $\pm$ 0.07 & 8.89 $\pm$ 0.16 \\
        
        13 & SSTGLMA G345.4077-00.9857 & 17:09:44.33 & -41:37:06.98 & 13.08 $\pm$ 0.07 & 12.12 $\pm$ 0.09 & 10.95 $\pm$ 0.13 & - \\
        
        14 & SSTGLMA G345.3787-01.0036 & 17:09:43.42 & -41:39:09.23 & 14.86 $\pm$ 0.20 & 12.71 $\pm$ 0.12 & 11.40 $\pm$ 0.12 & 10.44 $\pm$ 0.14 \\
        
        15 & SSTGLMA G345.3737-00.9888 & 17:09:38.63 & -41:38:52.06 & 8.85 $\pm$ 0.05 & 7.37 $\pm$ 0.05 & 5.98 $\pm$ 0.03 & 5.08 $\pm$ 0.03 \\
        
        16 & SSTGLMA G345.3817-01.0017 & 17:09:43.51 & -41:38:56.73 & 12.31 $\pm$ 0.05 & 11.16 $\pm$ 0.07 & 10.38 $\pm$ 0.07 & 10.38 $\pm$ 0.08 \\
        
        17 & SSTGLMA G345.4001-00.9274 & 17:09:27.85 & -41:35:24.13 & 12.68 $\pm$ 0.27 & 11.39 $\pm$ 0.15 & 10.49 $\pm$ 0.27 & - \\
        
        18 & SSTGLMA G345.4108-00.9542 & 17:09:36.79 & -41:35:50.55 & 10.22 $\pm$ 0.17 & 8.04 $\pm$ 0.15 & 6.61 $\pm$ 0.14 & - \\
        
        19 & SSTGLMA G345.4185-00.9139 & 17:09:27.89 & -41:34:02.03 & 12.28 $\pm$ 0.11 & 11.54 $\pm$ 0.13 & 10.52 $\pm$ 0.17 & - \\
        
        20 & SSTGLMA G345.4234-00.9164 & 17:09:29.46 & -41:33:53.09 & 11.95 $\pm$ 0.15 & 11.15 $\pm$ 0.15 & 10.10 $\pm$ 0.21 & - \\
        
        21 & SSTGLMA G345.4188-00.9600 & 17:09:39.82 & -41:35:39.95 & 9.71 $\pm$ 0.11 & 8.19 $\pm$ 0.06 & 6.73 $\pm$ 0.09 & 5.86 $\pm$ 0.15 \\
        
        22 & SSTGLMA G345.4419-00.9400 & 17:09:39.07 & -41:33:50.17 & 12.30 $\pm$ 0.09 & 10.93 $\pm$ 0.09 & 9.25 $\pm$ 0.12 & 8.12 $\pm$ 0.22 \\
        
        23 & SSTGLMA G345.4203-00.9622 & 17:09:40.69 & -41:35:40.41 & 11.99 $\pm$ 0.18 & 10.95 $\pm$ 0.28 & 8.53 $\pm$ 0.20 & - \\
        
        24 & SSTGLMA G345.4403-00.9771 & 17:09:48.34 & -41:35:14.45 & 13.83 $\pm$ 0.31 & 11.75 $\pm$ 0.14 & 10.21 $\pm$ 0.15 & - \\
        
        25 & SSTGLMA G345.3071-00.9269 & 17:09:09.94 & -41:39:51.86 & 15.07 $\pm$ 0.18 & 13.95 $\pm$ 0.15 & 12.41 $\pm$ 0.26 & - \\
        
        26 & SSTGLMA G345.3085-00.9257 & 17:09:09.89 & -41:39:45.32 & 14.62 $\pm$ 0.16 & 13.76 $\pm$ 0.13 & 12.59 $\pm$ 0.31 & - \\
        
        27 & SSTGLMA G345.3913-00.9156 & 17:09:23.13 & -41:35:24.08 & 12.16 $\pm$ 0.10 & 9.90 $\pm$ 0.06 & 8.35 $\pm$ 0.05 & 7.54 $\pm$ 0.10 \\
        
        28 & SSTGLMA G345.3931-00.9171 & 17:09:23.86 & -41:35:22.02 & 11.73 $\pm$ 0.20 & 10.95 $\pm$ 0.16 & 9.79 $\pm$ 0.29 & - \\
        
        29 & SSTGLMA G345.3592-00.9110 & 17:09:15.79 & -41:36:47.25 & 12.24 $\pm$ 0.11 & 11.29 $\pm$ 0.11 & 10.33 $\pm$ 0.20 & - \\
        
        30 & SSTGLMA G345.3436-00.9111 & 17:09:12.85 & -41:37:32.42 & 13.34 $\pm$ 0.12 & 12.22 $\pm$ 0.16 & 11.27 $\pm$ 0.16 & - \\
        
        31 & SSTGLMA G345.4901-00.8584 & 17:09:27.24 & -41:28:36.03 & 12.40 $\pm$ 0.07 & 11.41 $\pm$ 0.07 & 10.65 $\pm$ 0.06 & 10.14 $\pm$ 0.05 \\
        
        32 & SSTGLMA G345.4936-00.8511 & 17:09:26.04 & -41:28:10.27 & 14.27 $\pm$ 0.12 & 12.89 $\pm$ 0.11 & 11.73 $\pm$ 0.12 & 11.23 $\pm$ 0.10 \\
        
        33 & SSTGLMA G345.4568-00.8246 & 17:09:12.21 & -41:28:59.68 & 14.03 $\pm$ 0.10 & 13.15 $\pm$ 0.11 & 12.42 $\pm$ 0.27 & - \\
        
        34 & SSTGLMA G345.3860-00.8134 & 17:08:55.80 & -41:32:00.38 & 12.30 $\pm$ 0.06 & 11.03 $\pm$ 0.06 & 9.91 $\pm$ 0.05 & 8.43 $\pm$ 0.02 \\
        \hline \hline
	\end{tabular}
\end{table*}


 \begin{figure*}
 	\includegraphics[width=0.8\textwidth]{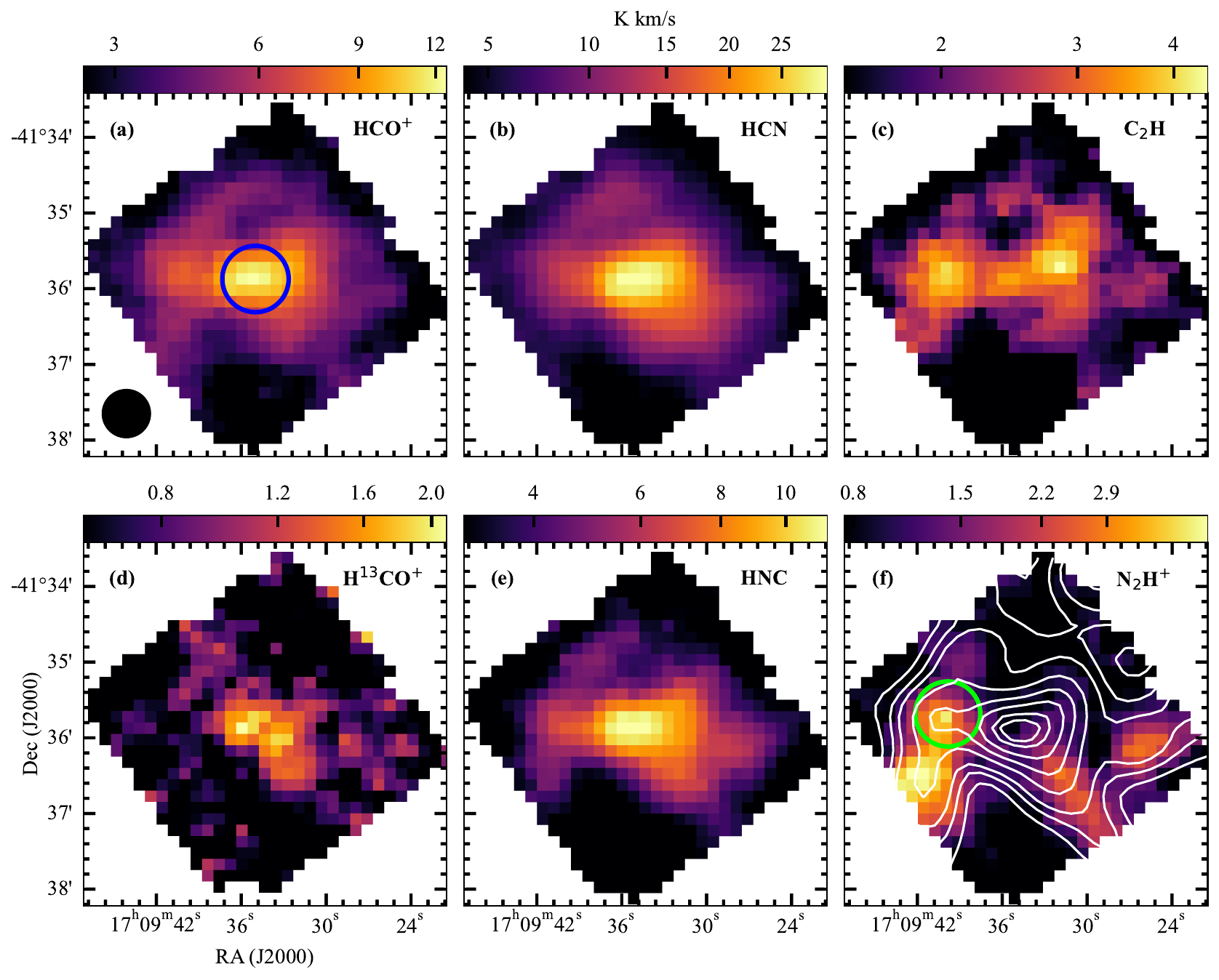}
 	\caption{Integrated intensity maps for molecules detected towards RCW~117 using the MALT90 survey. The beam is shown towards the bottom left in (a). The blue and green circles on (a) and (f) represent the region over which spectra are extracted and shown in Fig.~\ref{fig:spectra}. The contour levels in (f) represent the column density convolved to the \textit{MALT90} with levels $3 \times 10^{22}$, $4 \times4 10^{22}$, $5 \times 10^{22}$, $7 \times 10^{22}$, $9 \times 10^{22}$, $11 \times 10^{22}$, $13 \times 10^{22}$ and $14 \times 10^{22}$~cm$^{-2}$.}
 \label{fig:malt90mom0}
 \end{figure*} 



\bsp	
\label{lastpage}
\end{document}